\documentclass[a4paper,11pt]{article}
\usepackage{jcappub}

\newcommand{\figurewidth}{0.8\textwidth}
\newcommand{\gridfigurewidth}{\textwidth}

\abstract{Curvature perturbations induce gravitational waves (GWs) at second order, contributing to the stochastic gravitational wave background.  The resulting gravitational wave spectrum is sensitive to the evolutionary history of the universe and can be substantially enhanced by early matter-dominated (eMD) epochs, particularly if they end rapidly.  Such epochs can be caused by primordial black holes (PBHs) and non-topological solitons (Q-balls), for example.  Prior analysis approximated the end of the eMD epoch as instantaneous or used a Gaussian smoothing. In this work, we present a complete analysis fully incorporating their time-evolving decay rates.  We demonstrate that the resulting signal spectra from PBH, thin wall Q-ball, thick wall Q-ball, and delayed Q-ball eMD epochs are distinguishable for monochromatic distributions.  We then consider log-normal mass distributions and discuss the distinguishability of the various GW spectra. Importantly, we find that the resulting spectra from different types of matter, which decay at different rates, can be distinguished from the GW spectra from broader mass distributions.}

\begin{document}

\title{
Using Gravitational Wave Signals to Disentangle Early Matter Dominated Epochs
}

\author[a]{Matthew Pearce}
\affiliation[a]{School of Physics and Astronomy, Monash University, Melbourne 3800 Victoria, Australia}
\emailAdd{matthew.pearce@monash.edu}

\author[b]{Lauren Pearce}
\affiliation[b]{Department of Physics and Astronomy, James Madison University, Harrisonburg VA USA 22801 }
\emailAdd{lpearce@psu.edu}

\author[c]{Graham White}
\affiliation[c]{School of Physics and Astronomy, University of Southampton, Southampton SO17 1BJ, United Kingdom}
\emailAdd{g.a.white@soton.ac.uk}

\author[a]{Csaba Bal\'azs}
\emailAdd{csaba.balazs@monash.edu}

\date{\today}

\maketitle

\section{Introduction}

Upcoming gravitational wave (GW) experiments are expected to probe or constrain the stochastic GW background~\cite{Kosowsky:1992rz,Caprini:2019egz,Vilenkin:1981zs,Dunsky:2021tih}, which motivates study of potential contributions to this stochastic background.  One contribution is the GWs induced by curvature perturbations at second order~\cite{Kohri:2018awv,Espinosa:2018eve}.  Because these can be generated by primordial curvature perturbations, including those produced by inflation, the signal strength of this contribution is sensitive to the cosmological history of the universe.

In particular, during a matter-dominated epoch, the gravitational potential which controls these waves does not decay, even on sub-horizon scales, in contrast to its behavior during radiation domination~\cite{Kohri:2018awv,Espinosa:2018eve}.  Thus an early matter-dominated (eMD) epoch enhances the metric perturbations sourced by curvature perturbations.  Currently, there are no constraints on the existence or duration of such an epoch, as long as it ended prior to Big Bang Nucleosynthesis~\cite{Amin:2014eta,Hasegawa:2019jsa}.

Early matter-dominated epochs are well-motivated, as they can occur if the energy density of the universe is dominated by a heavy quasi-stable objects or by a rapidly oscillating scalar field.  Motivation can come from model-building concerns (for example, moduli fields in string-inspired extension of the Standard Model~\cite{Coughlan:1983ci,deCarlos:1993wie,Banks:1993en,Banks:1995dt} or heavy dark sector particles~\cite{Pospelov:2007mp,Arkani-Hamed:2008hhe,Berlin:2016vnh,Zhang:2015era,Erickcek:2020wzd,Erickcek:2021fsu}).  They can also be motivated phenomenologically by noting that both primordial black holes (PBHs) and non-topological solitons (Q-balls), which arise in a broad class of models, can cause eMD epochs ~\cite{Inomata:pbh,White:2021hwi,Kasuya:2022cko,Papanikolaou:2022chm}.  In this work, we focus on these scenarios.

Because such a wide variety of models lead to eMD epochs, it is natural to ask whether it is possible to determine the underlying physics from the resulting GW signal.  This is particularly important as it has been suggested that these GW signals have the potential to probe specific particle-physics models such as high scale supersymmetry~\cite{Flores:2023dgp}.

However, in the approximation of an instantaneous end to the eMD epoch, the resulting signal depends only on the horizon scale at the start of matter domination and its length~\cite{Inomata:sudden}, which does not allow for disentangling different physical scenarios, and accounting for heavy matter distribution functions via Gaussian smoothing as in~\cite{Inomata:pbh} does not alter this conclusion.

However, the GW signal depends sensitively on the speed at which the eMD epoch ends~\cite{Inomata:sudden,Inomata:gradual}.  In the fast transition limit, there is a resonant-like enhancement (known as the poltergeist mechanism~\cite{Inomata:pbh}), but in a slow transition, this resonance is replaced by a suppression~\cite{Inomata:gradual}.  Although both PBHs and Q-balls have rapid decays, their decay rates are not identical, and therefore a careful analysis which incorporates the finite transition time between matter and radiation domination can potentially disentangle these signals.  This paper does such an analysis, building on recent work~\cite{Pearce:2023kxp} which, for the first time, accurately calculates the GW spectrum in between the fast and slow limits.  

Our paper is arranged as follows.  In section~\ref{sec:GW_review} we review the production mechanism for GWs, detailing the evolution of perturbations in the synchronous gauge and providing integrals for the GW power spectra.  In section~\ref{sec:decay_rates} we then review the decay mechanisms of PBHs and Q-balls, motivating a time dependent form of the decay rate that encompasses both PBHs and various limiting Q-ball scenarios.  We then present our results for monochromatic mass distributions in section~\ref{sec:mono_results}, which clearly demonstrate the different signals from PBHs and different classes of Q-balls.  In section~\ref{sec:distributions} we then detail how our analysis can be modified to included extended mass distributions.  The results of this analysis follow in section~\ref{sec:results_mass}, where we see that the signals generally remain meaningfully different until they approach the severely suppressed slow transition limit.  Our conclusions are then presented in section~\ref{sec:conclusion}.

\section{Review of Gravitational Wave Signal}
\label{sec:GW_review}

First, we review the power spectrum of the GWs induced from the curvature perturbations, following~\cite{Ananda:2006af,Kohri:2018awv,Espinosa:2018eve,Inomata:2016rbd, Ananda:2006af,Baumann:2007zm}. Throughout the paper, we will use primes to denote differentiation with respect to conformal time $\eta = \int dt/a(t)$.

When considering the evolution of the perturbations, we work in the synchronous gauge, in which observers are comoving with the decaying matter, allowing for convenient evaluation of the decay rate.  In this gauge, the perturbed metric is
\begin{equation}
    ds^2 = a^2\left[-d\eta^2+\left(\delta_{ij}+H_{ij}+\frac{h_{ij}}{2}\right)dx^idx^j\right],
\end{equation}
where $a$ is the scale factor of the background FLRW metric, $h_{ij}$ are the tensor perturbations induced at second order , and the Fourier transform of the scalar part of the metric is given by
\begin{equation}
    H_{ij}=\hat{k}_i\hat{k}_j\gamma+\left(\hat{k}_i\hat{k}_j-\frac{1}{3}\delta_{ij}\right)6\epsilon.
\end{equation}
In this expression $\hat{\boldsymbol{k}}=\boldsymbol{k}/k$ and we have adopted the notation from reference~\cite{Inomata:pbh}, such that the fields $\gamma$ and $\epsilon$ correspond to $h$ and $\eta$, respectively, in references~\cite{Ma:1995ey,Poulin:2016nat}.

When we come to evaluating the GW power spectrum, the analysis has traditionally been done in the conformal Newtonian gauge.  In this gauge, the perturbed metric takes the form
\begin{equation}
\label{eq:metric}
    ds^2 = a^2\left[-(1+2\Phi)d\eta^2+\left((1-2\Psi)\delta_{ij}+\frac{h_{ij}}{2}\right)dx^idx^j\right],
\end{equation}
where $\Phi$ and $\Psi$ are the first order scalar perturbations, and $h_{ij}$ are again the second order tensor perturbations.  We will neglect anisotropic stress, and thus $\Psi$ is equal to the gravitational potential $\Phi$.  Note again that we use the notation of reference~\cite{Inomata:gradual}, with $\Phi$ and $\Psi$ corresponding to $\psi$ and $\phi$, respectively, in references~\cite{Ma:1995ey,Poulin:2016nat}.

The coupled Friedmann and continuity equations control the evolution of the matter energy density $\rho_m$, the radiation energy density $\rho_r$ and the scale factor $a$~\cite{Poulin:2016nat}
\begin{align}
\label{eq:fried_m}
    \rho_m'&=-(3\mathcal{H}+a\Gamma(\eta))\rho_m,\\
\label{eq:fried_r}
    \rho_r'&=-4\mathcal{H}\rho_r+a\Gamma(\eta)\rho_m,\\
\label{eq:fried_a}
    a'&=\frac{1}{\sqrt{3}M_{\rm Pl}}\sqrt{(\rho_m+\rho_r)}a^2,
\end{align}
where $M_{\rm Pl}=1/\sqrt{8\pi G}$ is the reduced Planck mass and we have noted that the decay rate will depend on conformal time.  The evolution of the Friedmann equations in our code is described in appendix~\ref{ap:background}, where we factor out redshifting and normalise all quantities to their value at $\eta_{\rm eq,1}$, the matter-radiation equality time at the start of the eMD epoch.  We measure all times in units of $\eta_{\rm eq,1}$ and denote the matter-radiation equality time at the end of the eMD epoch as $\eta_{\rm eq,2}$.

By considering the evolution of the phase space distribution of the decaying matter as in~\cite{Poulin:2016nat,Ma:1995ey}, the synchronous gauge equations of motion for the perturbations can be derived.  Defining energy density perturbations $\delta=\delta\rho/\rho$ and velocity divergences $\theta$, the matter and radiation perturbations satisfy
\begin{align}
	\label{eq:deltamsync}
    \delta_m^{(S)'}&=-\frac{\gamma'}{2}-\theta_m^{(S)}, \\
    \label{eq:deltarsync}
    \delta_r^{(S)'}&=-\frac{4}{3}\left(\theta_r^{(S)}+\frac{\gamma'}{2}\right)+a\Gamma(\eta)\frac{\rho_m}{\rho_r}(\delta_m^{(S)}-\delta_r^{(S)}),\\
    \label{eq:thetarsync}
    \theta_r^{(S)'}&=\frac{k^2}{4}\delta_r^{(S)}-a\Gamma(\eta)\frac{3\rho_m}{4\rho_r}\left(\frac{4}{3}\theta_r^{(S)}-\theta_m^{(S)}\right),
\end{align}
where the comoving coordinates keep $\theta_m^{(S)}=0$.  These are to be solved alongside the metric perturbations, which satisfy
\begin{align}
	k^2\epsilon-\frac{1}{2}\mathcal{H}\gamma'&=-\frac{3}{2}\mathcal{H}^2\left(\frac{\rho_m\delta_m^{(S)}+\rho_r\delta_r^{(S)}}{\rho_{\rm tot}}\right),\\
	k^2\epsilon' &= 2\mathcal{H}^2\frac{\rho_r}{\rho_{\rm tot}}\theta_r^{(S)}.
\end{align}

Evolution equations for the perturbations in Newtonian gauge, using the decay rate evaluated in the synchronous gauge, were derived in Ref.~\cite{Pearce:2023kxp}.  However, in this work, we directly solve the synchronous gauge perturbation equations, using the adiabatic initial conditions during the initial radiation-dominated (RD) epoch
\begin{equation}
\label{eq:ICs}
	\delta_{r,0}=-\frac{2}{3}C(k\eta^2),\;\delta_{m,0}=\frac{3}{4}\delta_{r,0},\;\theta_{r,0}=-\frac{1}{18}C(k^4\eta^3),\;\gamma_0=C(k\eta)^2,\;\epsilon_0=C\left(2-\frac{1}{18}(k\eta)^2\right).
\end{equation}
Given these initial conditions, we note that eq.~\eqref{eq:deltamsync} has the trivial solution, $\delta_m^{(S)}=-\gamma/2$.  

We note that some black hole production mechanisms may also produce isocurvature perturbations, which can also induce gravitational waves~\cite{Papanikolaou:2020qtd,Domenech:2020ssp,Domenech:2023jve}. Similar mechanisms may also produce isocurvature perturbations when Q-balls are produced; if these have different spectra, then they can induce a different gravitational wave spectrum at second order.  In current analyses (e.g., \cite{Bhaumik:2022pil,Bhaumik:2022zdd}), this signal is assumed to be independent of the poltergeist signal that we explore in this work.  We therefore leave the study of such a signal for future work.

Once a solution in the synchronous gauge has been obtained, the metric potentials in the conformal Newtonian gauge are given by
\begin{align}
	\label{eq:GTphi}
	\Phi&=\mathcal{H}\alpha+\alpha', \\
	\label{eq:GTpsi}
	\Psi&=\epsilon-\mathcal{H}\alpha,
\end{align}
where $\alpha=(6\epsilon+\gamma)'/(2k^2)$ and in the regime of interest $\Psi = \Phi$.  We set $C=5/6$ which corresponds to taking the normalisation of RD epoch, superhorizon modes to be $\Phi_0=10/9$, so that modes entering the horizon during the eMD epoch are normalised\footnote{We note that this convention is consistent with eq.~\eqref{eq:source}.  Other references, such as ref.~\cite{Kohri:2018awv}, normalise $\Phi$ differently but also correspondingly alter the normalization of the source.} to $\Phi=1$.

The details of how the scalar perturbations induce GWs are encoded in the source function
\begin{align}
    f(u,v,k,\Bar{\eta})=\frac{3}{25(1+w)}&\Bigl(2(5+3w)\Phi(uk,\Bar{\eta})\Phi(vk,\Bar{\eta})\nonumber\\
    &+4\mathcal{H}^{-1}(\Phi'(uk,\Bar{\eta})\Phi(vk,\Bar{\eta})+\Phi(uk,\Bar{\eta})\Phi'(vk,\Bar{\eta}))\nonumber\\
    &+4\mathcal{H}^{-2}\Phi'(uk,\Bar{\eta})\Phi'(vk,\Bar{\eta})\Bigr),
    \label{eq:source}
\end{align}
where $w$ is the equation of state and ${\cal H}=a^\prime/a$ is the conformal Hubble parameter.

This source function determines the time-dependence of the GW spectrum, which is described by 
\begin{equation}
\label{eq:I}
    I(u,v,k,\eta)=k\int_0^{\eta} d\Bar{\eta}\;\frac{a(\Bar{\eta})}{a(\eta)}kG_k(\eta,\Bar{\eta})f(u,v,k,\Bar{\eta}),
\end{equation}
where the Green's function $G_k$ is the solution to the GW equation of motion
\begin{equation}
    \label{eq:greens}
    G''_k(\eta,\Bar{\eta})+\left(k^2-\frac{a''(\eta)}{a(\eta)}\right)G_k(\eta,\Bar{\eta})=\delta(\eta-\Bar{\eta}).
\end{equation}

This allows us to calculate the GW power spectrum, given by 
\begin{equation}
\label{eq:spectra}
    \overline{\mathcal{P}_h(\eta,k)}=4\int_{0}^{\infty}dv\int_{|1-v|}^{1+v}du\left(\frac{4v^2-(1+v^2-u^2)^2}{4vu}\right)^2\overline{I^2(u,v,k,\eta)}\mathcal{P}_{\zeta}(uk)\mathcal{P}_{\zeta}(vk),
\end{equation}
where we have assumed a Gaussian distribution for the curvature perturbations and the overline denotes an oscillation average over time.

As we are interested in the GWs sourced by the primordial curvature perturbations, we use 
\begin{equation}
\label{eq:curvature}
{\mathcal{P}} _\zeta (k) = \Theta (k_{\rm max} - k) A_s \left( \frac{k}{k_\ast} \right) ^{n_s-1},   
\end{equation}
for the power spectrum.  To compare to detector reaches, we will use $A_s=2.1\times 10^{-9}$ for the amplitude at the pivot scale, $n_s=0.97$ for the spectral tilt, and $k_\ast=0.05 {\rm Mpc}^{-1}$ for the pivot scale~\cite{Planck:2018vyg}, although we will also use a flat power spectrum to illustrate differences in GW spectral shapes for epochs with different matter components.

The cutoff $k_{\rm max}$ is used to restrict our analysis to the linear regime.  It can be estimated as the mode that reaches $\delta_m=1$ at the end of the eMD epoch.  Early analyses~\cite{Inomata:gradual,Inomata:sudden} used the horizon scale at the start of the eMD epoch or a conservative estimate of $k_{\rm max}=450/\eta_{\rm eq,2}$ (if the mode that entered at the start of the eMD epoch reached non-linear evolution).  Modes that entered prior to the start of the eMD epoch were neglected as they decay prior to the start of the matter domination.  However, as noted in~\cite{Inomata:pbh,Kumar:2024hsi}, modes that enter shortly before matter domination undergo little decay and their contribution to the resulting GW signal may be significant, particularly for short eMD epochs.

Given our established computational framework, it is a simple task to evolve a single mode over the duration of the eMD epoch.  Therefore, in this work, we improve upon the conservative estimates for the cut-off by searching through different $k$ modes to find the one for which $\delta_m=1$ at the end of the eMD epoch, which we denote as $k_{\rm max}$.  All modes for which $k>k_{\rm max}$ enter before this mode and hence will all have become non-linear by the end of the eMD epoch.  In using this estimation of $k_{\rm max}$ for the non-linear cut-off, we are able to more completely consider the effect of the length of the eMD epoch on the spectrum, in conjunction with how rapid the transition to radiation domination is.  For all of the results presented in this paper, we have confirmed that the mode with $k=k_{\mathrm{max}}$ enters during the RD epoch that precedes the eMD one; that is, all of the modes that enter during matter domination evolve linearly through to the end of that epoch.  These modes which enter during the eMD epoch are the most affected by the matter-dominated epoch and its conclusion, and thus we choose parameters such that these can be studied with linear perturbation theory, as in prior work (e.g.,~\cite{Inomata:pbh, Pearce:2023kxp}).

Finally, from the power spectrum, we can derive the GW abundance
\begin{align}
    \Omega_{\rm GW}(\eta,k)&=\frac{\rho_{\rm GW}(\eta,k)}{\rho_{\rm tot}(\eta)}\nonumber\\
    &=\frac{1}{24}\left(\frac{k}{\mathcal{H}(\eta)}\right)^2 \overline{\mathcal{P}_h(\eta,k)}.
\end{align}
Following the eMD epoch, the gravitational potential decays and the GW spectrum will eventually become constant, at a time which we will denote as $\eta_c$.  The amplitude of the spectrum at the current time $\eta_0$ can then be found from the subsequent evolution through the standard RD epoch~\cite{Inomata:sudden}
\begin{equation}
\label{eq:GW_dilution}
	\Omega_{\rm GW}(\eta_0,k)=0.39\left(\frac{g_*(\eta_c)}{106.75}\right)^{-1/3}\Omega_{r,0}\Omega_{\rm GW}(\eta_c,k),
\end{equation}
where $g_*$ is the effective number of relativistic degrees of freedom and $\Omega_{r,0}$ is the current radiation energy density parameter.  In addition, to account for redshifting, we need to know the horizon scale at the end of the transition, which we obtain from the reheating time~\cite{Inomata:sudden}
\begin{equation}
\label{eq:reheat_time}
	\frac{\eta_R}{10^{-14}\textrm{Mpc}} =\left(\frac{g_{*s}}{106.75}\right)^{1/3} \left(\frac{g_*}{106.75}\right)^{-1/2}\left(\frac{T_R}{1.2\times10^7\textrm{GeV}}\right)^{-1},
\end{equation}
where $g_{*s}$ is the effective number of relativistic degrees of freedom of entropy and both $g_*$ and $g_{*s}$ are now evaluated at the reheating time.  In the above $T_R$ is the temperature the universe reheats to following the evaporation of matter.  The above is derived by assuming energy and entropy conservation for an instantaneous transition from matter to radiation domination.  We also assume that $\eta_R\simeq \eta_{\rm eq,2}$, which is only strictly valid for rapid transitions.  In cases where the transition is gradual and cannot be assumed to be instantaneous, the above assumptions would only result in a small error in the redshifting and additionally, we will see that these types of signals are well beyond the limits of future GW observatories.

This signal has been calculated in a variety of contexts (see Ref.~\cite{Domenech:2021ztg} for a review), including cosmological phase transitions~\cite{Barir:2022kzo} and early matter domination epochs~\cite{Assadullahi:2009nf,Alabidi:2013lya,Kohri:2018awv,Espinosa:2018eve,Inomata:sudden,Inomata:gradual}.  The latter is motivated by the observation that during matter domination the gravitational potential does not decay (as it does during radiation domination), even on sub-horizon scales, and this enhances the source $f$ given by equation \eqref{eq:source}.  The signal is further enhanced if the matter-dominated epoch ends rapidly~\cite{Inomata:sudden}; in this case, the matter overdensities are transformed into sound waves, which can resonantly enhance GW modes moving at a particular angle with respect to the incoming mode~\cite{Ananda:2006af,Inomata:pbh,White:2021hwi}.  

Early analyses considered the limit of an instantaneous transition between matter and radiation domination~\cite{Inomata:sudden}; recent work has included the effect of a finite decay time~\cite{Pearce:2023kxp}.  However, in order to study the transition between fast and slow decay rates, this work used a tanh profile for the decay width, which is not motivated by any particular physical model.  In the next section, we will replace the tanh profile with appropriate decay rates for PBH and Q-ball models.  Here, we note that to calculate the GW signal we follow~\cite{Pearce:2023kxp}, which fully incorporates the transition period between matter and radiation domination, avoiding all step-function approximations.\footnote{In particular, we use the same code and so the technical details of Appendix B of~\cite{Pearce:2023kxp} apply to this work as well.  Besides incorporating physically motivated decay rates, the only significant difference is that we work entirely in the synchronous gauge until we calculate the gravitational potential using eq.~\eqref{eq:GTpsi}, whereas~\cite{Pearce:2023kxp} transforms the evolution equations to the Newtonian gauge.}

\section{Decay Rates of Physically-Motivated Matter Components}
\label{sec:decay_rates}

Both primoridial black holes and non-topological solitons have been argued to produce matter-dominated epochs that end rapidly, leading to an enhanced GW signal~\cite{Inomata:pbh,White:2021hwi}.  In both cases, the effective decay rates have been argued to increase as the objects become smaller, leading to rapid transitions.  In this section, we consider the different decay rates of Q-balls and PBHs, determining the appropriate $\Gamma(\eta)$ expressions to use in the synchronous evolution equations (equations \eqref{eq:deltamsync}, \eqref{eq:deltarsync} and \eqref{eq:thetarsync} above).  This will allow us to apply the analysis of~\cite{Pearce:2023kxp} to calculate the differing signal profiles accurately.

As a PBH decays, its mass evolves as~\cite{Inomata:pbh}
\begin{align}
 \dfrac{dM_{\mathrm{PBH}}}{dt} &= - \dfrac{A}{M_{\mathrm{PBH}}^2}.
\label{eq:PBHdmdt}
\end{align}
The coefficient is
\begin{align}
A &= \dfrac{\pi \mathcal{G} g_{H*} M_{\rm Pl}^4}{480},
\end{align}
where $\mathcal{G} \approx 3.8$ is a gray-body factor and the spin-weighted decays of freedom in the Hawking radiation $g_{H*}$ depends weakly on temperature, but we neglect this here.  This equation can be solved analytically to find 
\begin{align}
M_{\mathrm{PBH}} &= \left( M_{\mathrm{PBH},i}^3  - A t \right)^{1 \slash 3}.
\end{align}
It is convenient to define the evaporation time through $M_{\mathrm{PBH},i} = \left( A t_{\mathrm{eva}} \right)^{1 \slash 3}$, resulting in $M_{\mathrm{PBH}} = \left( A (t_{\mathrm{eva}} - t) \right)^{ 1 \slash 3}$.  The effective decay rate for a single black hole is then 
\begin{align}
\Gamma_{\rm PBH} &= - \dfrac{1}{M_{\mathrm{PBH}}} \dfrac{dM_{\mathrm{PBH}}}{dt} = \dfrac{1}{3 (t_{\mathrm{eva}} - t)}.
\label{eq:gamma_pbh}
\end{align}

Next we consider non-topological solitons (Q-balls).  These are extended scalar field configurations whose (quasi-)stability is ensured by a conserved charge.  Specifically, the energy of the Q-ball field configuration is less than the mass energy of the equivalent number of free charged quanta.  Unlike black holes, there are a variety of Q-balls types distinguished by the profile of the scalar vacuum expectation value (VEV).  In our analysis, we will consider thin wall, thick wall, and delayed Q-balls.

Thin wall (Coleman) Q-balls are characterized by a constant VEV $v$ throughout the interior, which drops to zero in a thin shell~\cite{Coleman:1985ki}.  These states have energy per unit charge $\omega = \sqrt{2 V(v) \slash v^2}$, which is set by the Q-ball potential $V$.  Any potential in which $V(\phi) \slash \phi^2$ is minimized at nonzero $\phi = v$ has Q-ball states; one particularly well-motivated potential can be generated via gravity mediated supersymmetry breaking~\cite{Coleman:1985ki}.

Although a Q-ball composed of field $\phi$ cannot decay into individual $\phi$ quanta, it is not necessarily completely stable.  If the conserved charge is carried by other fields, it may be possible for the Q-ball to decay if the quanta of these fields are sufficiently light.  We describe the decay rate of an individual $\phi$ quanta with the coupling $y$ between the scalar and decay products,
\begin{align}
\Gamma_{\phi } = \dfrac{3 \omega y^2}{8 \pi}.
\end{align}
If the decay could happen throughout the Q-ball, the decay rate of the Q-ball would be $Q \Gamma_\phi$ times this.  However, in this case the Q-ball's decay rate per unit charge is constant and therefore the effective decay rate does not increase as the Q-ball decays.  Thus, we do not expect a sufficiently rapid transition from matter to radiation domination to observe the resonant enhancement.

However, as was considered in~\cite{White:2021hwi}, decays may be kinematically forbidden in the interior of the Q-ball, as the large VEV may contribute to the mass of the other fields involved in the interaction.  Alternatively, if the decay products are fermions, decays in the interior of the Q-ball may be forbidden due to Fermi blocking if the decay products cannot efficiently diffuse out of the Q-ball.

If the decay is only allowed for radii $r > R_*$, we would expect the decay rate for the Q-ball to be 
\begin{align}
\Gamma_{\mathrm{Q-ball}} &= \dfrac{dQ}{dt} = \dfrac{3 Q  \omega y^2}{8 \pi} \dfrac{R^3 - R_*^3}{R^3}.
\end{align}

If decays in the interior of the Q-ball are forbidden due to Fermi blocking, then the decays occur in a shell, of width $ \sim (yv)^{-1}$, in which diffusion is efficient.  Thus $R - R_* \approx 1 \slash y v$, where we assume that $1 \slash (yv) \ll R$. Thus,
\begin{align}
\dfrac{dQ}{dt \, dA} \approx \frac{9 Q y \omega }{32 \pi ^2 R^3 v}
=\frac{3  y v\omega ^2}{8 \pi }
\end{align}
using $R = \left( \dfrac{3}{4\pi} \dfrac{Q}{\omega v^2} \right)^{1 \slash 3}$ for thin wall Q-balls~\cite{Coleman:1985ki}.  A more careful calculation~\cite{Cohen:1986ct} improves the numerical coefficient 
\begin{align}
\dfrac{dQ}{dt \, dA} &= \dfrac{y v \omega^2}{64 \pi},
\label{eq:dQdtdA-fermi}
\end{align}
although as with the black holes, the specific numerical coefficient turns out to unimportant to the GW signal as it is absorbed into the evaporation time.

We next discuss kinematic blocking.  We assume the decay products acquire a mass $\sim g v$ within the Q-ball and thus decays are forbidden if $2 g v > \omega$.  For a thin wall Q-ball, decays will happen only in the thin shell of width $\sim (gv)^{-1}$ in which the VEV nears zero, leading to 
\begin{align}
\dfrac{dQ}{dt \, dA} = \frac{3  y^2 v \omega ^2}{8 \pi  g},
\label{eq:dQdtdA-kin}
\end{align}
where we note that the numerical factor may vary slightly if e.g.\ only one decay product has a large mass or the decay products have different couplings $g$ controlling their masses.

Using the thin wall Q-ball radius, both expressions lead to 
\begin{align}
\dfrac{dQ}{dt} &= - \mathcal{C} Q^{2 \slash 3}
\label{eqref:thindQdt}
\end{align}
for a constant $\mathcal{C}$, where 
\begin{align}
\mathcal{C}_{\mathrm{Fermi}} &=  \dfrac{1}{32} \left( \dfrac{9}{2 \pi^2} \right)^{1 \slash 3}
\dfrac{y \omega^{4 \slash 3}}{v^{1 \slash 3}} 
,\qquad 
\mathcal{C}_{\mathrm{kin}}  = \dfrac{3}{4} \left( \dfrac{9}{2 \pi^2} \right)^{1 \slash 3}
\dfrac{y^2 \omega^{4 \slash 3}}{g v^{1 \slash 3}} 
\end{align}
for Fermi and  kinematic blocking respectively.

For a thin wall Q-ball with constant $v$ and $\omega$, the effective decay rate is 
\begin{align}
\Gamma_{\rm thin} &= - \dfrac{1}{M} \dfrac{dM}{dt} = - \dfrac{1}{Q} \dfrac{dQ}{dt} 
= \dfrac{3}{t_{\mathrm{eva}} - t }
\label{eq:Gamma_thinwall}
\end{align}
where we have integrated to find $Q(t)$ and the evaporation time is set by the constant $\mathcal{C}$ and the initial charge through 
\begin{align}
Q_i &=  \left( \mathcal{C} \cdot t _{\mathrm{eva}} \right)^3.
\end{align}
We have also used the fact that for a thin wall Q-ball, $M = Q \omega$ is a linear relation.

We note that eq.~\eqref{eqref:thindQdt} implies $\Gamma_{\mathrm{thin}} \propto Q^{-1 \slash 3} \sim M^{-1 \slash 3}$, which has been used to argue the transition effectively speeds up.  However, this speed up is slower than PBHs, which have $\Gamma_{\mathrm{PBH}} \sim M^{-3}$, as can be seen using eq.~\eqref{eq:PBHdmdt}.  We will see below that because it is the product with the energy density, $\rho_M \Gamma$, that matters, the thin wall spectrum will actually be qualitatively similar to the slow transition limit.

For completeness we observe that if decays occur throughout the volume of the Q-ball, we would have $dQ \slash dt \propto Q$ leading to $\ln(Q_0 \slash Q) = A t$.  Because the charge asymptotically approaches zero, the evaporation time is ill-defined.

Not all Q-balls are thin wall Q-balls; in fact, the thin wall regime must always break down at sufficiently small charges.  At sufficiently small charges, the Q-ball approaches the thick wall regime~\cite{Kusenko:1997ad}.  The charge at which the thick wall approximation becomes valid is typically $~\mathcal{O}(100)$, which is many, many orders of magnitude smaller than the initial charge~\cite{Kusenko:1997ad,Pearce:2012jp}.  At even smaller charges, quantum corrections become important~\cite{Graham:2001hr,Tranberg:2013cka}.  Because these Q-balls are in the final stages of their evaporation, the deviation from the thin wall behavior will not be important.  Therefore in the ``thin wall'' results presented below we use decay rate~\eqref{eq:Gamma_thinwall} until the Q-balls entirely evaporate at $t_{\mathrm{eva}}$.  We note that departures from the semi-classical regime are also expected for black holes~\cite{Dvali:2011aa,Dvali:2012en}.  In some scenarios, this could suppress the resulting signal~\cite{Bhaumik:2024qzd}; however, we do not incorporate such a suppression here due to the significant uncertainties.

Thick wall Q-balls, however, exist beyond the small-charge regime of thin wall Q-balls.  For example, the Q-balls in supersymmetric models with gauge-mediated SUSY breaking are thick wall Q-balls, even at large values of the charge~\cite{Multamaki:1999an}; specifically, for a potential of the form 
\begin{align}
V(\Phi) &= m^4 \log \left( 1 + \dfrac{\Phi^2}{m^2} \right) + \dfrac{\Phi^6}{\Lambda^2}
\end{align}
in which the last term is negligible. The VEV inside the non-topological soliton is no longer constant, varying with distance as 
 \begin{align}
\phi(r) &= m Q^{1 \slash 4} \cdot \dfrac{  \sin \left( \sqrt{2} m \pi Q^{-1 \slash 4} r \right) }{ \sqrt{2} m \pi Q^{-1 \slash 4} r }
\end{align}
out to a maximum radius
\begin{align}
R &= \dfrac{\pi} {\sqrt{2} m \pi Q^{-1 \slash 4}} = \dfrac{1}{\sqrt{2} m} Q^{1 \slash 4} .
\end{align}
Similarly the energy per unit charge now depends on the total charge, 
\begin{align}
\omega &=  \sqrt{2} m \pi Q^{-1 \slash 4}.
\end{align}

As above, we expect the effective decay rate to increase if the decays are forbidden within the volume of the Q-ball and occur only in a shell near its surface, due to the surface-to-volume ratio which is small for large Q-balls but increases as the Q-ball decays and shrinks.  Again we consider both kinematic and Fermi blocking, although as above they will share the same qualitative behavior.

As above, we take $g \phi(r)$ to be the effective mass of the decay products within the Q-ball and note that the width of the shell is set by $g \phi(r) > \omega \slash 2$.  We will assume that this is satisfied for $r \geq R - \epsilon$ with $\epsilon \ll R $ .  Expanding for small $\epsilon$, 
\begin{align}
\epsilon = \frac{\pi }{2 g m \sqrt[4]{Q}}
\end{align}
and $\epsilon \ll R$ if $\pi \slash (\sqrt{2Q} g) \ll 1 $ which is satisfied for sufficiently large charge (As for thin wall Q-balls above, we assume the regime in which this is violated is  negligible as the Q-ball is nearly fully evaporated).  Repeating the analysis as above, we find 
\begin{align}
\dfrac{dQ}{dt \, dA} = -\frac{9 m^3 y^2}{16 g \sqrt[4]{Q}}
\end{align}
after expanding in $\epsilon$ and keeping the linear term.  As above, $y$ is the coupling of the scalar to the decay product.

We can also consider the case where the fermions cannot diffuse out of the Q-ball efficiently.  In the thin wall case, the decays happen only in a shell of width $(yv)^{-1}$.  The situation is more complicated in the thick wall case in which the VEV varies, but we will approximate the thickness as $(y m Q^{1 \slash 4})^{-1}$, and we will note that any numerical factors will be absorbed into the evaporation time definition below.  Note that consistency requires $1 \slash y \phi_0  \ll R $, which is equivalent to $\sqrt{2} \slash \sqrt{Q} y \ll 1$.  This gives 
\begin{align}
\dfrac{dQ}{dt \, dA} = -\frac{9 m^3 y}{8 \pi  \sqrt[4]{Q}}
\end{align}
keeping the term linear in the shell width $(1 \slash y \phi_0 )$.

After multiplying by the area of the shell, we find both scenarios are described by 
\begin{align}
\dfrac{dQ}{dt} &= - \mathcal{C} Q^{1 \slash 4}
\end{align}
where, up to the numerical factors neglected above, 
\begin{align}
\mathcal{C}_{\mathrm{Fermi}} = 
\frac{9}{4}  y  m, \qquad 
\mathcal{C}_{\mathrm{kin}} = 
\frac{9 \pi}{8}  \dfrac{y^2m }{ g}    .
\end{align}
This equation can be integrated to derive a relation connecting the initial charge to the evaporation time, 
\begin{align}
Q_i &= \left(  \mathcal{C} t _{\mathrm{eva}} \right)^{4 \slash 3}.
\end{align}
Because the mass of the Q-ball is $M = \omega Q = \sqrt{2} m \pi Q^{3 \slash 4} $, the effective decay rate is 
\begin{align}
\Gamma_{\mathrm{thick}} &= - \dfrac{1}{M} \dfrac{dM}{dt} 
= - \dfrac{3}{4} \cdot \dfrac{1}{Q} \dfrac{dQ}{dt}
= \dfrac{1}{(t_{\mathrm{eva}} - t) },
\end{align}
where again integration was used to find $Q(t)$.

Finally, we note that the GW signal has been estimated from so-called delayed Q-balls in the instantaneous transition limit in~\cite{Kasuya:2022cko,Kawasaki:2023rfx}.  The charge of these Q-balls evolves as 
\begin{align}
Q(t) = Q_i \left( 1 - \dfrac{t}{t_{\mathrm{eva}}} \right)^{4 \slash 5}
\end{align}
where the evaporation time is related to the initial charge by  $t_{\mathrm{eva}} = 4 Q_i^{5 \slash 4} \slash 5 \mathcal{C} $ where $\mathcal{C}$ is a constant ultimately determined by the potential and decay couplings.  As the mass of the delayed Q-ball scales as $Q^{ 3 \slash 4}$, these have an effective decay rate
\begin{align}
\Gamma_{\mathrm{delayed}} &= - \dfrac{1}{M} \dfrac{dM}{dt} 
= \dfrac{3}{5(t_{\mathrm{eva}} - t) }.
\end{align}

These are the four physically-motivated decay rates we consider in this work; they all can be described using 
\begin{align}
\Gamma(t) = \dfrac{n}{t_{\mathrm{eva}} - t}
\label{eq:gendecayrate}
\end{align}
where the constants $n$ are given in Table \ref{tab:eff_decay_rate_n}.

\begin{table}
    \centering
    \begin{tabular}{cc}
     \textbf{ Model }   &  $n$\\
\hline      PBHs   & $1 \slash 3$\\
\hline      Thin-wall Q-balls (surface decay)  & $3$ \\
\hline      Thick-wall Q-balls (surface decay)   & $1$ \\
\hline      Delayed Q-balls   & $3 \slash 5$\\
 \hline   \end{tabular}
    \caption{The coefficient that appears in the effective decay rate \eqref{eq:gendecayrate} for the models considered in this work.}
    \label{tab:eff_decay_rate_n}
\end{table}

For monochromatic mass distributions, these decay widths can be substituted directly into the background equations \eqref{eq:fried_m}, \eqref{eq:fried_r}, and \eqref{eq:fried_a}.  Technical details of how we solved these are given in Appendix \ref{ap:background}.  To find the GW signal, we then also use this decay rate in the first order perturbation equations derived above.

\section{Results for Monochromatic Spectra}
\label{sec:mono_results}

In this section, we present the results of our analysis for monochromatic mass spectra, fully accounting for the fact that the transition between the eMD epoch and the RD epoch proceeds at different rates for different types of matter.  This goes beyond the instantenous approximation that has been used to estimate the signal in the literature, and importantly, introduces the possibility of distinguishing between the signals.

We first discuss qualitative features of the GW spectra to determine whether it is theoretically possible to determine the type of matter that caused the eMD epoch before proceeding to consider whether upcoming experiments can sufficiently probe the GW spectrum for this to be practical.  To study the shape of the GW spectrum we first consider a flat curvature spectrum as the GW signal then depends only on $k \eta_{\mathrm{eq}}$, allowing for easy comparison, although when we consider the observational reach of upcoming experiments we use Eq.~\eqref{eq:curvature}.

\begin{figure}[tbp]
    \centering
    \includegraphics[width=\figurewidth]{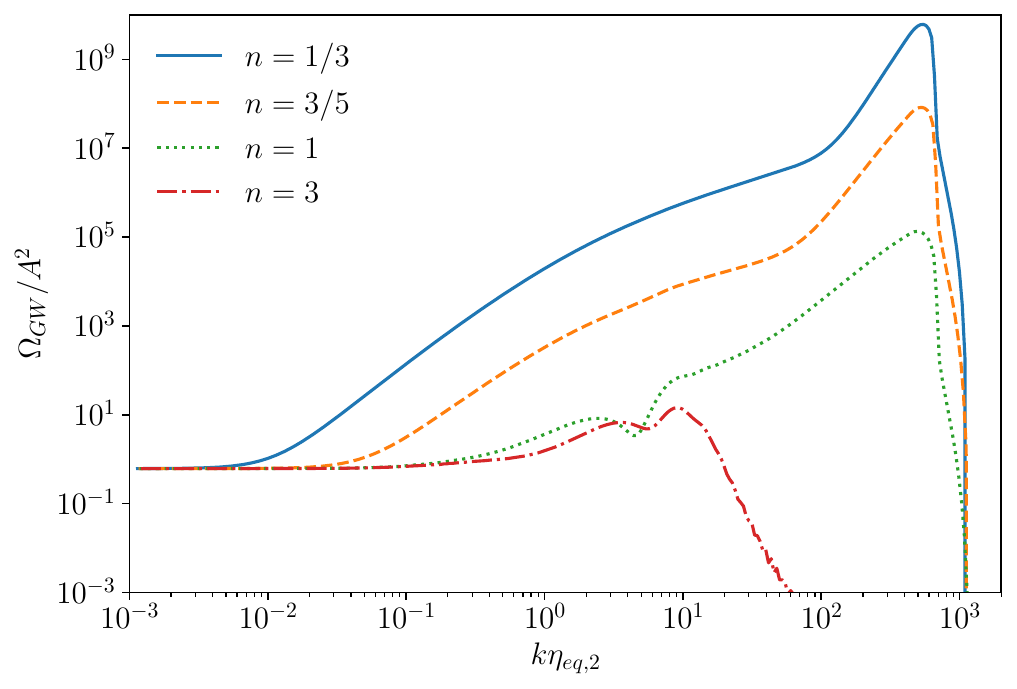}
    \caption{A comparison of the GW power spectrum originating from a eMD epoch dominated by PBH (solid-blue curve, with $n=1/3$), delayed Q-ball (orange-dashed curve, with $n=3/5$), thick-walled Q-ball (dotted-green curve, with $n=1$) and thin-walled Q-ball (dot-dashed-red curve, with $n=3$).  For all situations considered, we have set the length of the eMD epoch to be $\eta_{\rm eq,2}/\eta_{\rm eq,1}=500$.}
\label{fig:mono_results2}
\end{figure}

Fig.~\ref{fig:mono_results2} shows the resulting signal for all four scenarios on a single plot for an eMD epoch of fixed length.  We see that as expected, the signal is greatest for PBHs, which have the smallest $n$ and thus the fastest transition, and it is severely suppressed for thin-wall Q-balls which have the largest $n$ and thus the slowest transition between matter domination and radiation domination.   In fact, this transition is sufficiently slow that the poltergeist peak is severely suppressed and replaced with the suppression previously observed for time-independent decay rates~\cite{Inomata:gradual}.

The extreme suppression for the thin-wall Q-balls may be surprising, in light of the arguments of \cite{White:2021hwi}.  As noted, it is true that for thin-wall Q-balls the charge decay rate $dQ/dt$ divided by $Q$ scales as a negative power of the charge, and so the decay rate is larger for lower charge Q-balls.  For thin wall Q-balls, the charge is proportional to the mass, and so as noted $\Gamma \sim M^{-1 \slash 3}$.  We note that in the background evolution equations, this is multiplied by the matter energy density, and so the quantity $\rho_m \Gamma$ scales as a positive power of the mass, in contrast to the other models.  Since it is $\rho_m \Gamma$ which ultimately controls the rate at which the scale factor changes, this accounts for the absence of the poltergeist enhancement.

Most importantly, the PBH, thick wall, and delayed Q-ball all demonstrate the poltergeist-type peak at $k \eta_{\mathrm{eq,2}} \sim 10^3$, although the signal strengths differ.  All of the signals asymptote to the same value at low values of $k\eta_{\mathrm{eq}}$ as expected.  Thus, if the length of the eMD epoch was known, a measurement of the peak amplitude is sufficient to identify the type of matter.  

\begin{figure}[tbp]
    \centering
    \includegraphics[width=\gridfigurewidth]{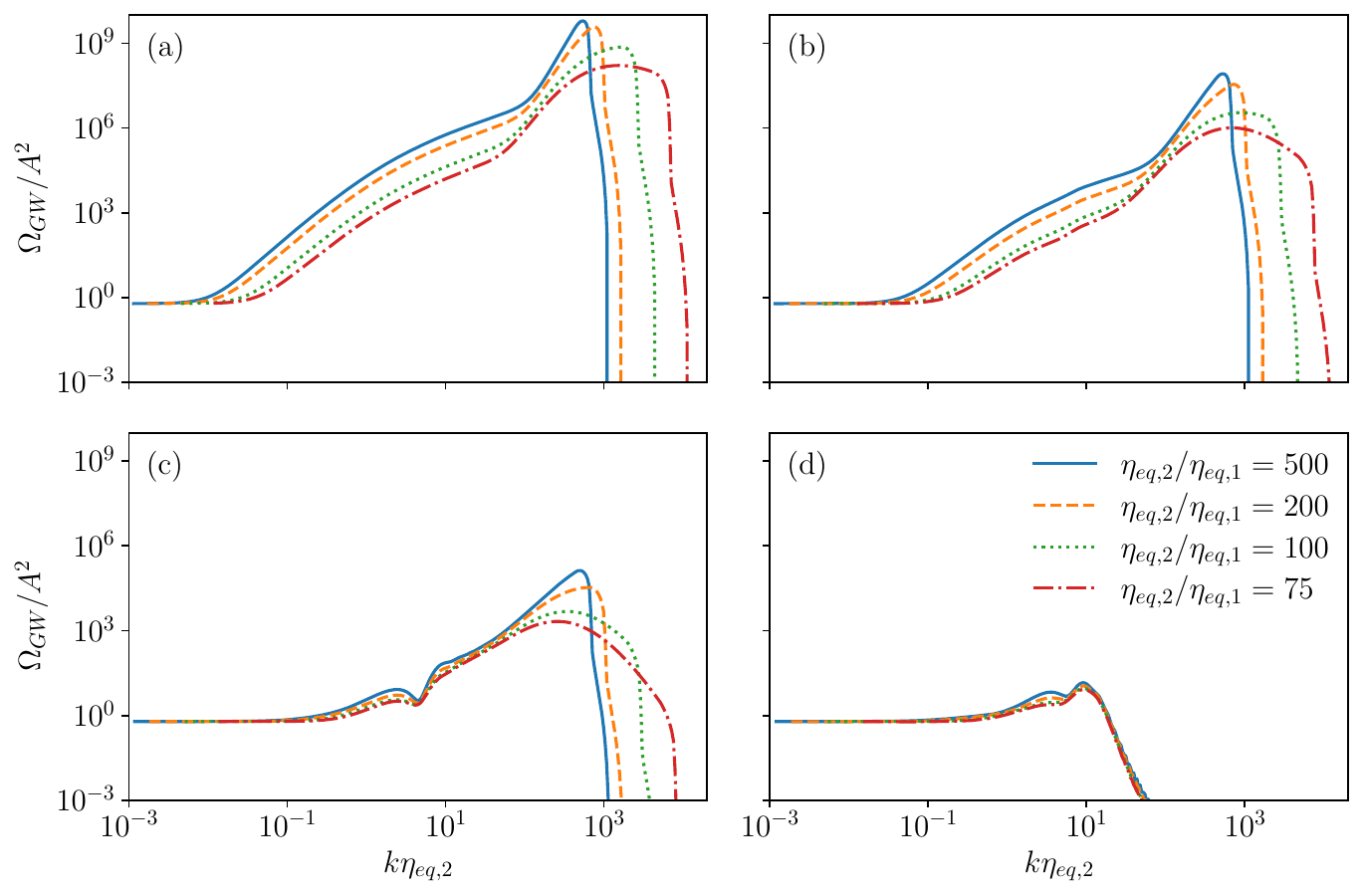}
    \caption{GW power spectrum induced from a scale invariant power spectrum ($n_s=1$) during a (a) PBH epoch, (b) delayed Q-ball epoch, (c) thick-walled Q-ball epoch and (d) a thin-walled Q-ball epoch.  The different curves correspond to eMD eras of different length, denoted by $\eta_{\rm eq,2}/\eta_{\rm eq,1}$.}
    \label{fig:mono_results1}
\end{figure}

However, the length of the eMD epoch may not be independently known.  Therefore, it is reasonable to study the evolution of these GW signal curves as the length of the eMD epoch varies.  The four plots in Fig.~\ref{fig:mono_results1} demonstrate the resulting GW signals for the four types of matter epochs described above for eMD epochs of different lengths.  We note that although we consider epochs of different lengths, for all of these, the mode that enters at the beginning of the eMD epoch is still undergoing linear evolution at its conclusion.

We see that the length of the eMD epoch significantly affects the shape of the GW signal, except in the thin-wall Q-ball scenario, in which the signal is highly suppressed as the transition is slow compared to other cases.  In particular, the signal is suppressed for shorter length eMD epochs as expected.  More importantly, the shape of the peak changes as it broadens for shorter epochs.  Therefore, by probing the shape of the peak one can determine the length of the eMD epoch, and as noted as above, the amplitude of the signal then allows for the determination of the matter's decay properties.

\begin{figure}[tbp]
    \centering
    \includegraphics[width=\figurewidth]{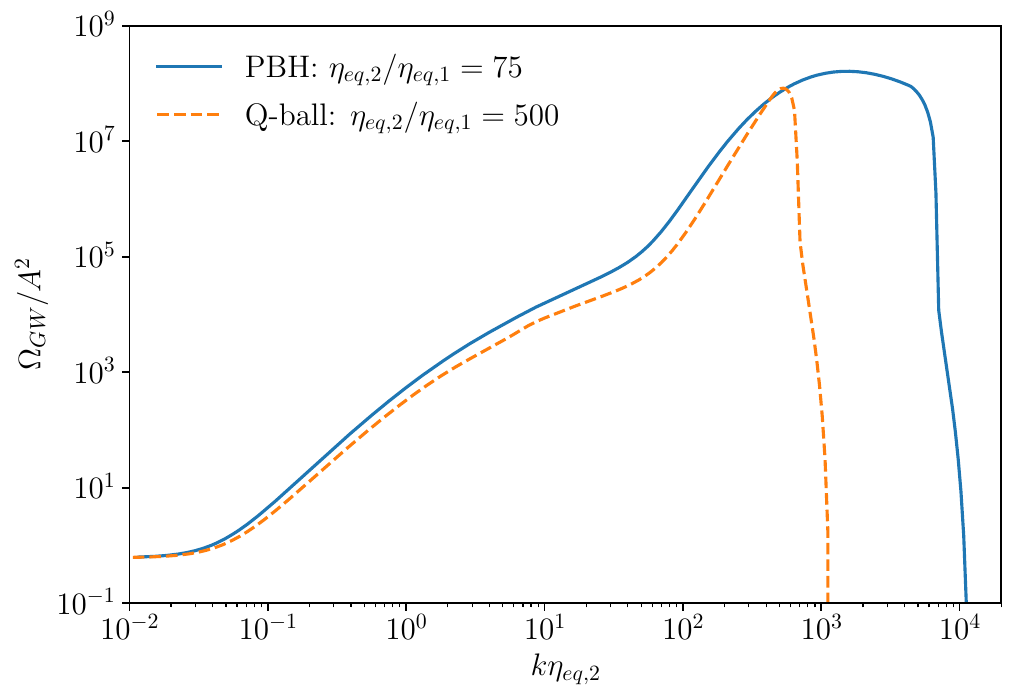}
    \caption{A comparison of the GW signal induced from a long eMD epoch dominated by delayed Q-balls, with $\eta_{\rm eq,2}/\eta_{\rm eq,1}=500$, to a short eMD epoch dominated by PBHs, with $\eta_{\rm eq,2}/\eta_{\rm eq,1}=75$.}
    \label{fig:mono_results_matched}
\end{figure}

The shift of the spectrum to larger values of $k\eta_{{\rm eq},2}$ is consistent with the results of reference~\cite{Inomata:pbh} but requires discussion.  For shorter eMD eras, the non-linear scale $k_{\rm max}$ becomes larger and thus modes with larger values of $k$ can be included in our analysis.  This is because for sufficiently short eMDs, all modes that enter during the eMD undergo linear evolution throughout the entire epoch, and even some modes which enter during the preceding RD epoch may remain non-linear if we account for the initial decay they experience.

Because the cutoff $k_{\mathrm{max}}$ comes from our inability to include modes which evolve non-linearly, the GW signal above the cutoff is unknown.  Thus, for example, the $\eta_{\mathrm{eq},2} \slash \eta_{\mathrm{eq},1} = 500$ signal is unknown for $k \eta_{\mathrm{eq}} > 10^3$, and we cannot say for certain how it compares to the $\eta_{\mathrm{eq},2} \slash \eta_{\mathrm{eq},1} = 75$ signal, which may be reliably computed.  Thus, we emphasize that the claim that one can determine the length of the eMD epoch through the shape of the GW signal is based specifically on the difference in signal shape for values of $k \eta_{\mathrm{eq}} \lesssim 10^3$, in which the signal can be reliably calculated for all four values of $\eta_{\mathrm{eq}}$ considered.

We note that although we leave detailed calculations for future work, we expect the signal in the GW signal for $k > k_{\mathrm{max}}$ to be suppressed.  As noted in~\cite{Inomata:pbh}, the $k\eta_{\mathrm{eq},2}$ value of the peak is set by the interplay between the enhancement of the spectrum from rapid oscillations of the gravitational potential and the evolution of large $k$ modes near the non-linearity bound at the end of the eMD epoch.    Note that for $\eta_{\mathrm{eq},2} \slash \eta_{\mathrm{eq},1} = 500$, the GW signal has dropped away significantly from the poltergeist peak before (to the immediate left of) the cutoff.  This follows from the fact that the modes that enter during the preceding RD epoch do not contribute significantly to the poltergeist signal.  

With this in mind, we consider Fig.~\ref{fig:mono_results_matched}, which shows a PBH signal and delayed Q-ball signal, where the lengths of the eMD epochs have been adjusted to produce signals of approximately the same amplitude.  Even restricting ourselves to consider the spectrum for  $k\eta_{\mathrm{eq},2} < 10^3$, set by $k_{\mathrm{max}}$ for the delayed Q-balls, we see that the PBH spectrum is broader than the Q-ball spectrum.  Thus a measurement of the amplitude and shape of the peak is sufficient, in theory, to determine both the length of the early matter epoch and the rate at which it ended.

A relatively unique feature of the slower transitions (thick-wall and thin-wall Q-balls) is the slight dip around $k \eta_{\mathrm{eq},2} = 10$.  A mild form of this dip can be seen in e.g.~Figures 3 and 4 of \cite{Pearce:2023kxp}.  The origin can be understood from \cite{Inomata:gradual}; at this point, there is a significant cancellation in the so-called cross-term contribution.  If this cross-term is dominant, the dip is evident in the overall signal.  See also Fig.~4 of \cite{Ananda:2006af}.

Finally, we note that these plots were made using a flat primordial curvature power spectrum, as the resulting signal is then an invariant function of $k \eta_{\mathrm{eq}}$.  However, CMB measurements suggest the spectral tilt $n_S = 0.97$ is different from one; accounting for this necessarily introduces a non-trivial scale dependence into the signal.  We include this in the plots in section \ref{sec:results_mass}, where we also include finite width mass distributions.

\section{Including Log-Normal Mass Distributions}
\label{sec:distributions}

The above analysis assume a monochromatic mass distributions for the PBHs and Q-balls.  Extended mass distributions generally lengthen the transition time between matter and radiation domination.  This  suppresses the signal, as the resonance effect decreases and, additionally, a subtle cancellation further suppresses the signal~\cite{White:2021hwi}.  This in turn makes the epochs harder to distinguish as for sufficiently broad mass distributions they all approach the slow limit.  It is thus desirable to study how the above GW spectra behave as we move away from the monochromatic limit.  

We will consider log-normal mass distributions, which are particular well-motivated for PBHs~\cite{Kannike:2017bxn}.  Q-balls have a variety of production mechanisms, but it has been argued that the Affleck-Dine fragmentation mechanism generically leads to log-normal distributions~\cite{Dolgov:2017nmh}. 

Finite-width distributions have been considered in Ref.~\cite{Inomata:pbh}; however, this work used Gaussian smoothing to approximate the signal instead of fully incorporating the finite transition period, as we now do.  In this section, we will describe how we incorporated mass distributions into our analysis before presenting our results in the next section.

We parameterise the initial mass distribution as 
\begin{align}
\label{eq:initial_distribution}
\rho_{m,i} &= \rho_{m,i} \int \beta(M_i) \, d \ln M_i,
\end{align}
where the distribution function is normalized such that $\int \beta(M_i) \, d \ln M_i = 1$ and the energy density at later times is given by 
\begin{align}
\rho_{m} (t)  &= \rho_{m,i} \int \beta(M_i,t) \, d \ln M_i.
\label{eq:dist}
\end{align}

The log-normal mass distribution is described by 
 \begin{align}
\beta_i = \beta(t=0) &= \dfrac{\mathcal{N}}{\sqrt{2\pi} \sigma } \exp \left(- \dfrac{\left( \ln(M_i \slash M_{0} ) \right)^2}{2 \sigma^2} \right) ,
\end{align}
where $M_{0}$ is a central Gaussian value (of the initial mass distribution), not an initial value of a particular black hole mass, and $\sigma$ describes the width of the Gaussian.  $\mathcal{N}$ is the normalization constant. As was noted in~\cite{Inomata:pbh}, the energy density at later times can be found by evolving each mass and including the dilution from spacetime expansion; this is confirmed in Appendix \ref{ap:background}.

We note that this can be converted into a distribution over evaporation times, as each initial mass corresponds to an evaporation time:
\begin{align}
M_{\mathrm{PBH},i} &= A^{1 \slash 3} t_{\mathrm{eva}}^{1 \slash 3}, \nonumber \\
M_{\mathrm{thin},i} &= \omega \mathcal{C}^3 t_{\mathrm{eva}}^3, \nonumber \\
M_{\mathrm{thick},i} &= \sqrt{2} m \pi Q^{3 \slash 4}_i = \sqrt{2} m \pi \mathcal{C} t_{\mathrm{eva}}, \nonumber \\
M_{\mathrm{delayed},i} &= \dfrac{4 \sqrt{2} \pi}{3} \zeta M_F (5 \mathcal{C} \slash 4)^{3 \slash 5} \cdot t_{\mathrm{eva}}^{3 \slash 5},
\end{align}
where for the Q-balls, we have used $M = \omega Q$, except for the delayed Q-ball, where the mass expression from \cite{Kasuya:2022cko} was used (and $\zeta$ and $M_F$ are model parameters).   We note the coefficients cancel and the initial distributions over evaporation times are 
 \begin{align}
\beta_{\mathrm{PBH},i}  &= \dfrac{\mathcal{N}_{\mathrm{PBH}}}{\sqrt{2\pi} \sigma } \exp \left(- \dfrac{\left( \ln(t_{\mathrm{eva}}^{1 \slash 3} \slash t_{\mathrm{eva},0}^{1 \slash 3} ) \right)^2}{2 \sigma^2} \right) , \nonumber \\
\beta_{\mathrm{thin},i}  &= \dfrac{\mathcal{N}_{\mathrm{thin}}}{\sqrt{2\pi} \sigma } \exp \left(- \dfrac{\left( \ln(t_{\mathrm{eva}}^3 \slash t_{\mathrm{eva},0}^3 ) \right)^2}{2 \sigma^2} \right) , \nonumber \\
\beta_{\mathrm{thick},i}  &= \dfrac{\mathcal{N}_{\mathrm{thick}}}{\sqrt{2\pi} \sigma } \exp \left(- \dfrac{\left( \ln(t_{\mathrm{eva}} \slash t_{\mathrm{eva},0} ) \right)^2}{2 \sigma^2} \right) , \nonumber \\
\beta_{\mathrm{delayed},i}  &= \dfrac{\mathcal{N}_{\mathrm{delayed}}}{\sqrt{2\pi} \sigma } \exp \left(- \dfrac{\left( \ln(t_{\mathrm{eva}}^{3 \slash 5} \slash t_{\mathrm{eva},0}^{3 \slash 5} ) \right)^2}{2 \sigma^2} \right) , 
\end{align}
where $t_{\mathrm{eva},0}$ is the evaporation time of the central Gaussian mass value.  These expressions are more convenient for our purposes as we have expressed the decay rates $\Gamma$ in terms of time $t$ and the evaporation time $t_{\mathrm{eva}}$ above.  For convenience we use the integration variable $\ln( t_{\mathrm{eva}} \slash t_{\mathrm{eva},0}) $, and thus the normalization constants are determined by the condition 
\begin{align}
    \int \beta_i \; d \ln(t_{\mathrm{eva}}\slash t_{\mathrm{eva},0}) &= 1.
\end{align}

In order to incorporate these distributions into our background and perturbation equations, we make the replacement
\begin{align} \Gamma(t) \rho_m  \rightarrow  \rho_{m,i} \int \Gamma(t_{\mathrm{eva}},t) \beta(t_{\mathrm{eva}},t) \, d\ln t_{\mathrm{eva}},
\label{eq:replacement}
\end{align}
as well as using eq.~\eqref{eq:dist} as appropriate.  This integral and the one in eq.~\eqref{eq:dist}, are non-trivial for the given log-normal mass distributions.  In Appendix~\ref{ap:Decay} we detail the careful approximate and numerical procedures we used to evaluate them accurately.

Having made these substitutions, we proceed to evaluate the background equations (see Appendix \ref{ap:background}) and the perturbation equations.  We emphasize that we fully incorporate the transition period between the radiation and matter dominated epochs, including in the Green's functions and the gravitational potential that determines the source function.  As was observed in Ref.~\cite{Pearce:2023kxp}, this is important in avoiding spurious features in the GW spectra.

\section{Results for Mass Distributions}
\label{sec:results_mass}

Here we present the results of our analysis, fully accounting for the fact that the transition between matter and radiation dominated epochs proceeds faster or slower for different types of matter as well as distributions of matter.  Above, we saw that monochromatic distributions of PBHs and different types of Q-balls produce different profiles.  However, we expect the signal to decrease and approach the slow transition limit as the distribution departs from the monochromatic limit, which means that we expect the GW signals to become more similar.  In this section, we explore qualitatively how this happens.  

\begin{figure}[tbp]
    \centering
    \includegraphics[width=\gridfigurewidth]{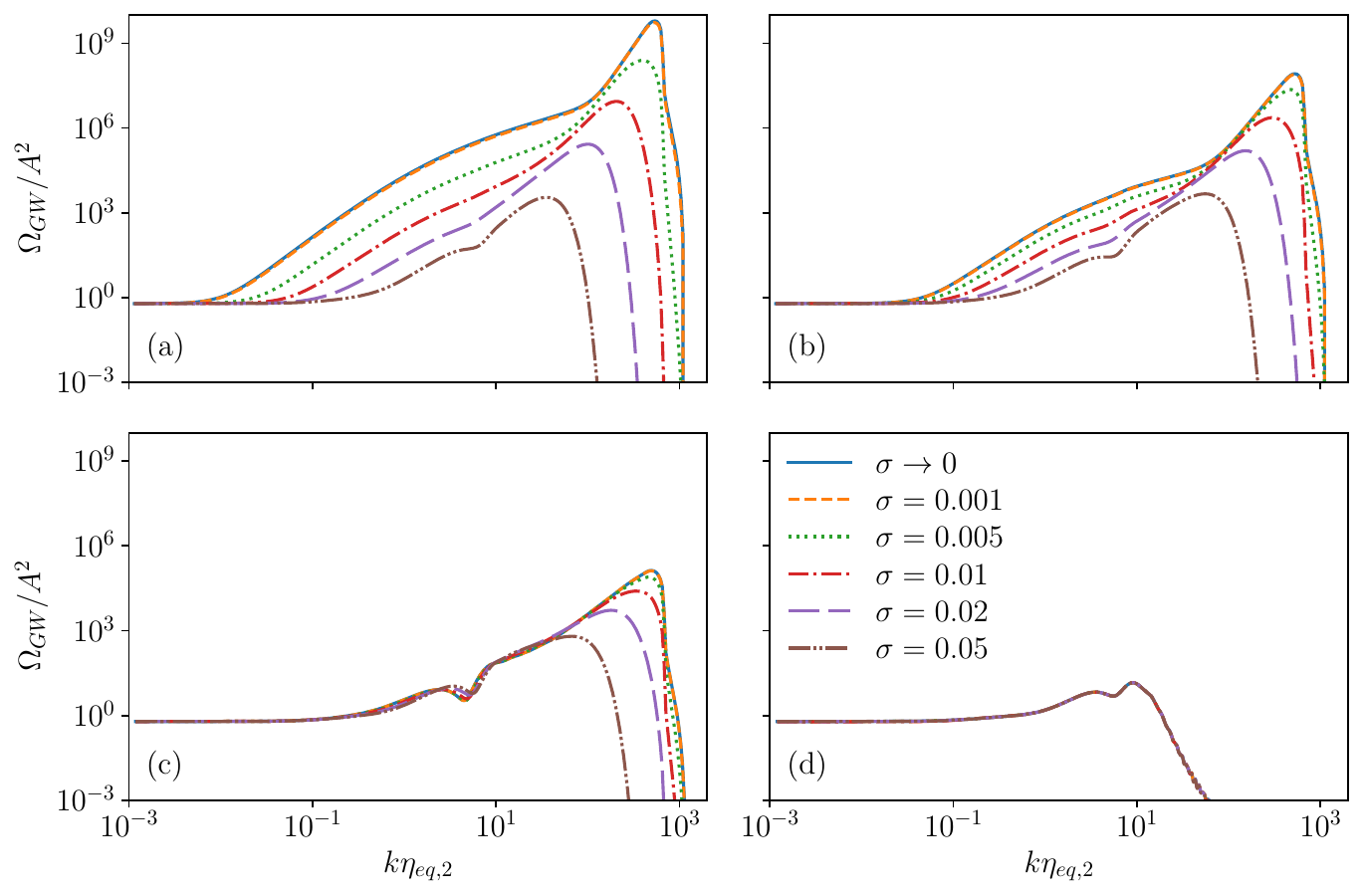}
    \caption{GW power spectrum induced from a scale invariant power spectrum ($n_s=1$) during a (a) PBH epoch, (b) delayed Q-ball epoch, (c) thick-walled Q-ball epoch and (d) a thin-walled Q-ball epoch.  In all cases the length of the eMD epoch is fixed to $\eta_{eq,2}/\eta_{eq,1}=500$ by adjusting the value of $t_{\rm eva,0}$.  The different curves correspond to different widths $\sigma$ for the initial mass distributions in eq.~\eqref{eq:initial_distribution}.}
    \label{fig:distribution_1}
\end{figure}

\begin{figure}[tbp]
    \centering
    \includegraphics[width=\gridfigurewidth]{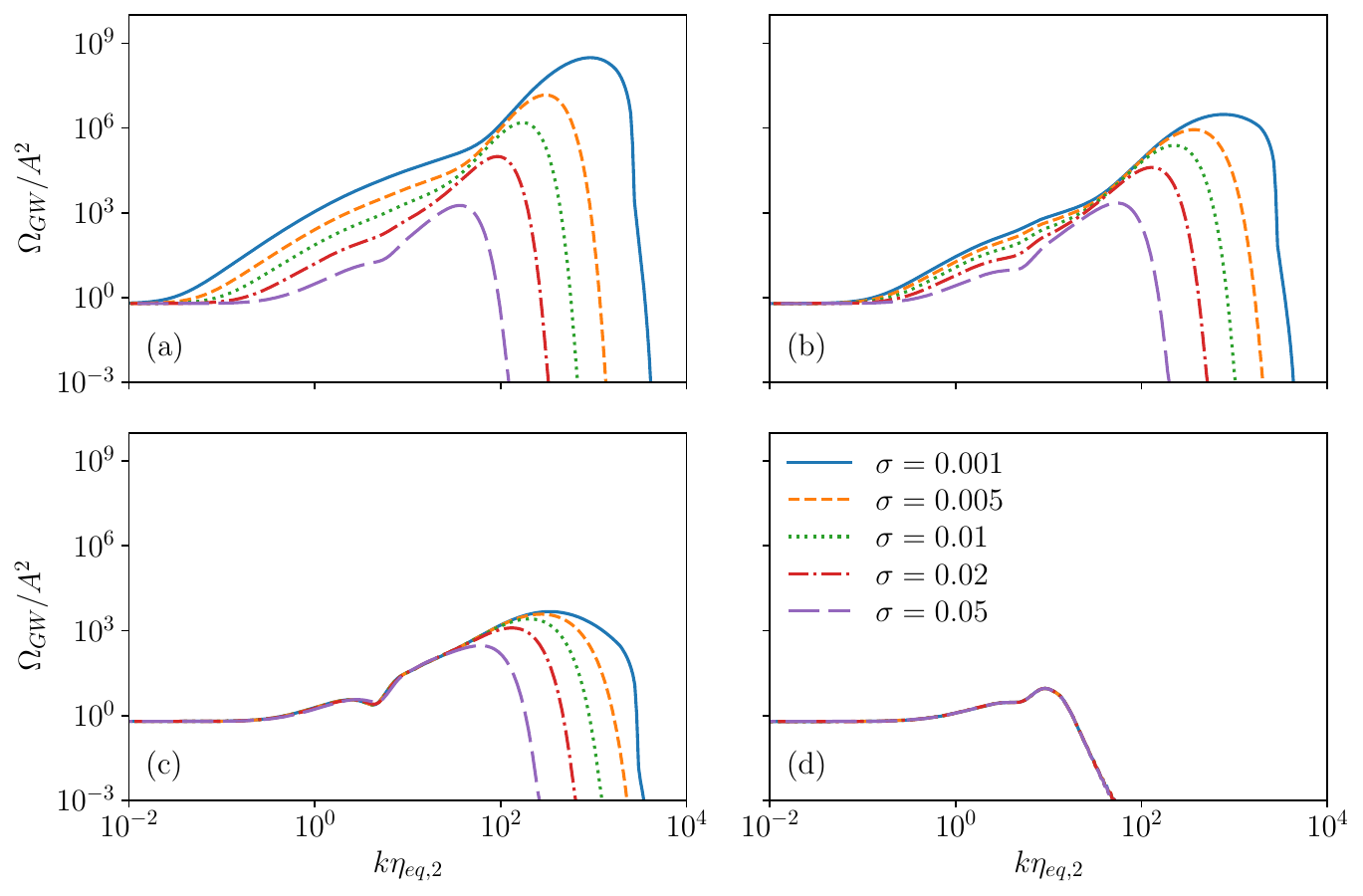}
    \caption{GW power spectrum induced from a scale invariant power spectrum ($n_s=1$) during a (a) PBH epoch, (b) delayed Q-ball epoch, (c) thick-walled Q-ball epoch and (d) a thin-walled Q-ball epoch.  In all cases the length of the eMD epoch is fixed to $\eta_{eq,2}/\eta_{eq,1}=100$ by adjusting the value of $t_{\rm eva,0}$.  The different curves correspond to different widths $\sigma$ for the initial mass distributions in eq.~\eqref{eq:initial_distribution}.}
    \label{fig:distribution_short}
\end{figure}

\begin{figure}[tbp]
    \centering
    \includegraphics[width=\gridfigurewidth]{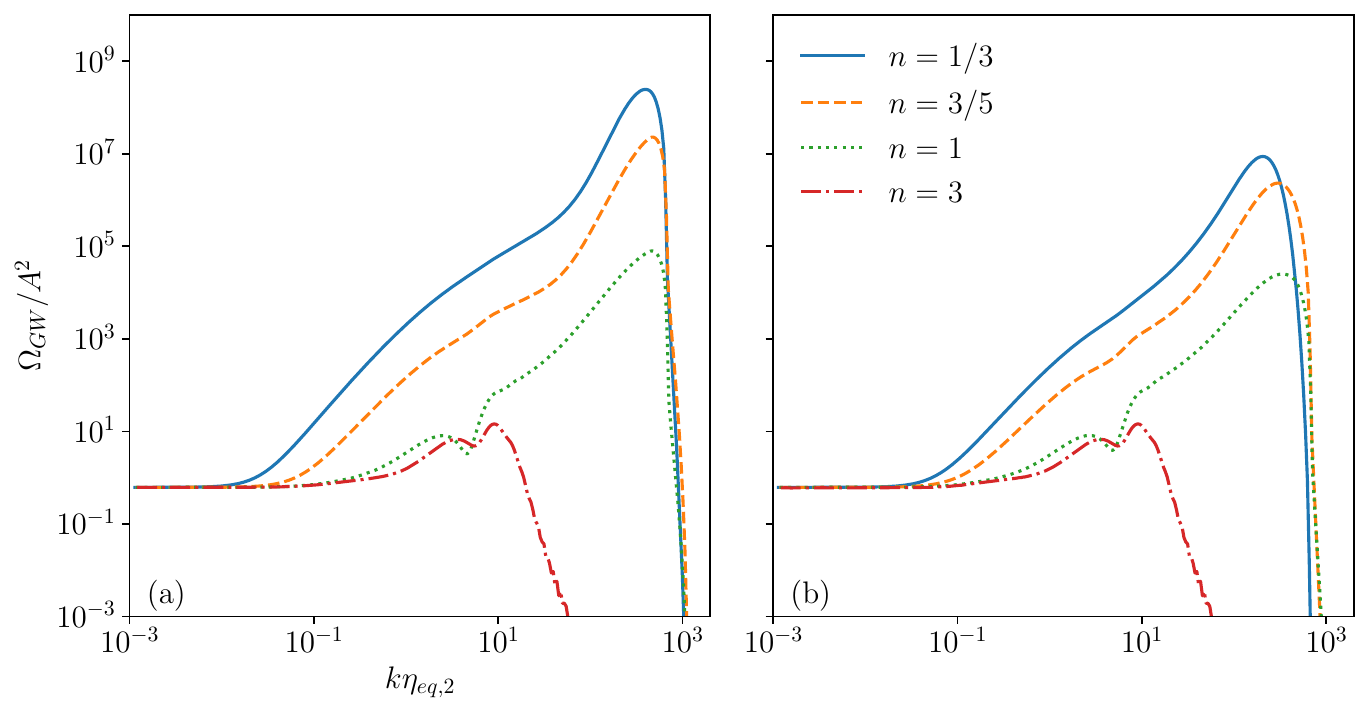}
    \caption{GW power spectrum induced from a scale invariant power spectrum ($n_s=1$) during an early eMD epoch with initial width (a) $\sigma=0.005$ and (b) $\sigma=0.01$.  In both cases the length of the eMD epoch is fixed to $\eta_{eq,2}/\eta_{eq,1}=500$.  The different curves correspond to the different types of matter occupying the universe as in table~\ref{tab:eff_decay_rate_n}.}
    \label{fig:distribution_sigma}
\end{figure}

\begin{figure}[tbp]
    \centering
    \includegraphics[width=\gridfigurewidth]{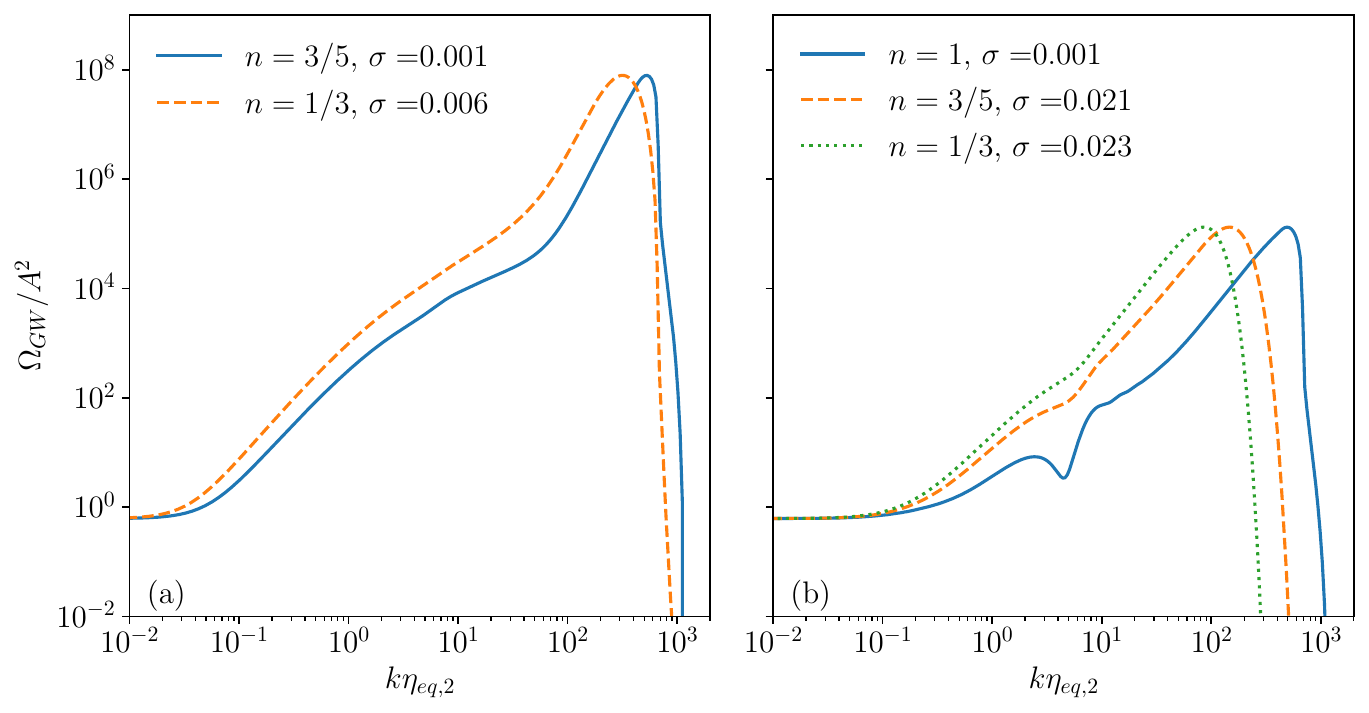}
    \caption{A comparison of the GW induced from different kinds of eMD eras where the widths of the initial distributions $\sigma$ have been adjusted such that the amplitude of the spectra matches that of an eMD epoch dominated by (a) delayed Q-balls ($n=3/5$) with $\sigma=0.001$ and (b) thick-walled Q-balls ($n=1$) with $\sigma=0.001$.  While adjusting $\sigma$, we simultaneously adjust $t_{\rm eva,0}$ so as to fix the length of the eMD epoch to $\eta_{eq,2}/\eta_{eq,1}=500$.}
    \label{fig:distribution_matched}
\end{figure}

We present our results for the different types of matter in Fig.~\ref{fig:distribution_1}.  First, we see that as the width of the Gaussian becomes small, the results approach the monochromatic limit presented above.  However, as the Gaussian widens, the signal is suppressed as expected, at least for the PBHs, thick wall, and delayed Q-balls.  For thin wall Q-balls, we see little impact from the Gaussian because the signal is already suppressed due to the comparatively slow decay.  Importantly, we see that the larger the monochromatic signal is, the narrower the Gaussian must be for the monochromatic limit to be applicable.  Thus, for PBHs, the monochromatic limit is applicable only for Gaussians with width $\sigma \lesssim 0.001$.  

Fig.~\ref{fig:distribution_short} shows the same set of plots for a shorter eMD epoch ($\eta_{\mathrm{eq},2} \slash \eta_{\mathrm{eq},1} = 100$).  As above, we see that the signal is suppressed and broadens.  We note that we do not see the GW spectrum from very broad mass distributions approaching the short eMD GW spectrum, because here we have adjusted $t_{\rm eva,0}$ to keep the length of the eMD epoch fixed.  This enables us to isolate the suppression that arises specifically from the gradual end of the eMD epoch.

In these plots, signals from the same matter models ($n$ value) are grouped together while $\sigma$ varies.  The inverse is shown in Fig.~\ref{fig:distribution_sigma}, where we fix the width of the distribution $\sigma$ and vary $n$.  We see that the signal curves are different, and particularly, the thin and thick wall Q-balls are easy to identify.  However, the PBH and delayed Q-balls signals are superficially similar, which leads to the concern that it may not be possible to disentangle these models if a positive signal is observed.  

This is explored further in Fig.~\ref{fig:distribution_matched}.  In this case, we have adjusted the width of the distribution so that the amplitudes of the GW spectrum peak match.  We see that the spectral shape is different, although it is not quite as pronounced as when we varied the length of the eMD epoch above.  Thus, we conclude that if a positive detection is made, it will be possible to distinguish e.g.\ the signal a narrow delayed Q-ball distribution from the signal from a PBH model which has been adjusted to have the same amplitude, either by shortening the eMD epoch or by broadening the mass distribution.

Importantly, we observe that widening the Gaussian distribution distorts the GW signal shape in a qualitatively different manner than the suppression from shorter eMDs which was explored in section \ref{sec:mono_results}; in particular, although the peak is suppressed, it does not significantly broaden.  Therefore, a sufficiently precise measurement of the GW spectrum can distinguish a shorter eMD epoch from a broader distribution of initial masses.

\begin{figure}[tbp]
	\centering
	\includegraphics[width=\figurewidth]{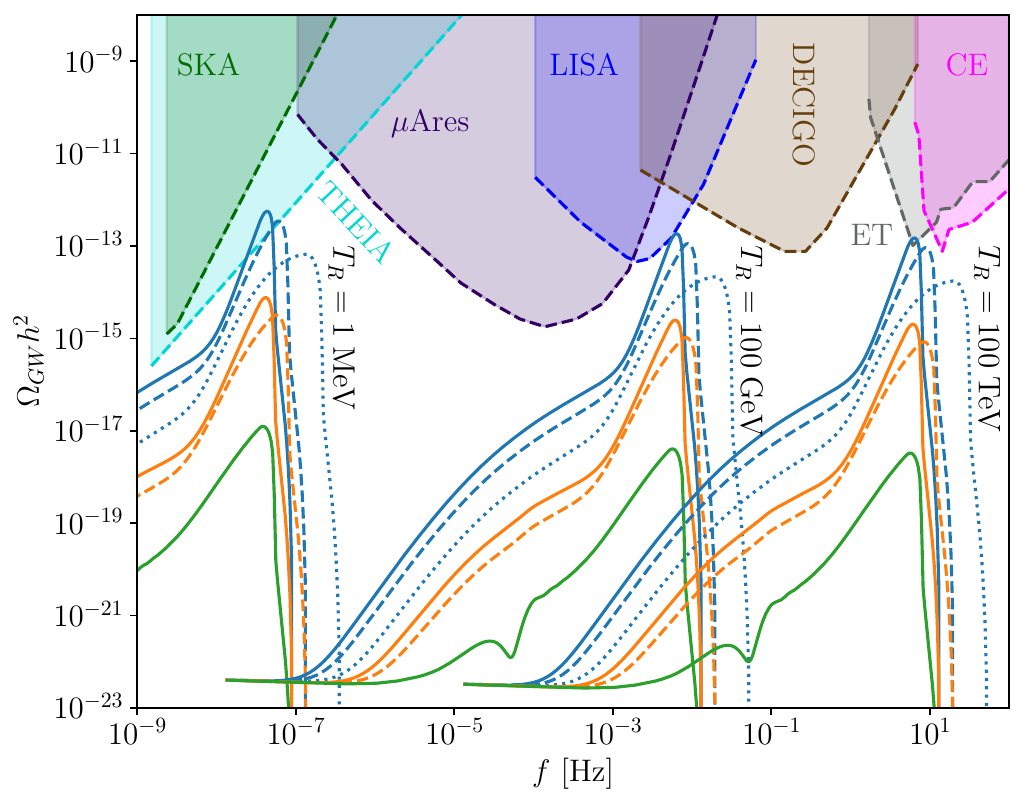}
	\caption{The GW spectrum induced from a curvature spectrum with tilt $n_s=0.97$ from transitions with reheating temperatures of $1\;{\rm MeV}$, $100\;{\rm GeV}$ and $100\;{\rm TeV}$.   In each scenario, we consider eMD eras dominated by PBHs (blue), delayed Q-balls (orange) and thick-wall Q-balls (green), with monochromatic mass distributions.  We consider eMD eras with the durations $\eta_{\rm eq,2}/\eta_{\rm eq,1}=500$ (solid curves), $\eta_{\rm eq,2}/\eta_{\rm eq,1}=225$ (dashed curves) and $\eta_{\rm eq,2}/\eta_{\rm eq,1}=100$ (dotted curves).  We contrast these to a selection of future GW experiments: LISA (blue) with an observing run of four years~\cite{Caprini:2019egz}, the Einstein Telescope (grey) for one year~\cite{Moore:2014lga,Maggiore:2019uih,Inomata:2018epa}, the Cosmic Explorer~\cite{Reitze:2019iox} (pink), DECIGO (brown) with three units for one year~\cite{Moore:2014lga,Kawamura:2020pcg}, $\mu$Ares (purple) for ten years~\cite{Sesana:2019vho}, THEIA (cyan) for twenty years~\cite{thetheiacollaboration2017theia,Garcia-Bellido:2021zgu} and SKA~\cite{Janssen:2014dka} (green).}
    \label{fig:Sensitivities_Monochromatic}
\end{figure}

\begin{figure}[tbp]
	\centering
	\includegraphics[width=\figurewidth]{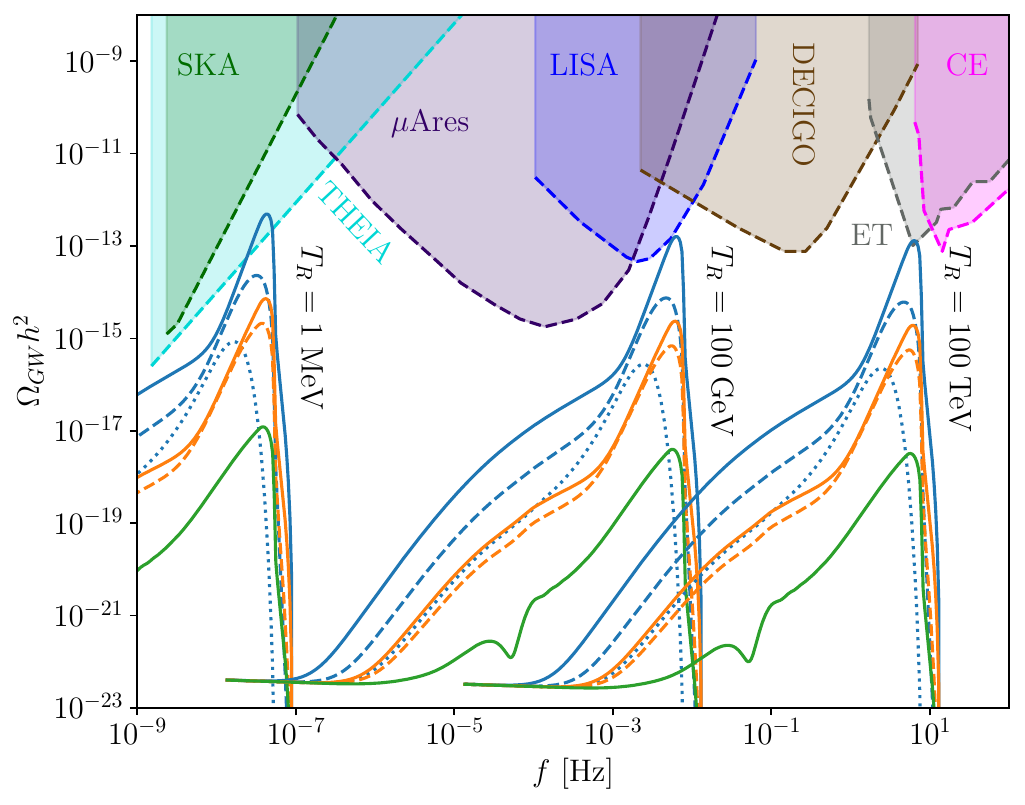}
	\caption{The GW spectrum induced from a curvature spectrum with tilt $n_s=0.97$ from transitions with reheating temperatures of $1\;{\rm MeV}$, $100\;{\rm GeV}$ and $100\;{\rm TeV}$.  In each scenario, we consider eMD eras dominated by PBHs (blue), delayed Q-balls (orange) and thick-wall Q-balls (green), for $\sigma = 0.001$ (solid curves), $\sigma = 0.005$ (dashed curves) and $\sigma = 0.01$ (dotted curves).  For all situations considered, we have set the length of the eMD epoch to be $\eta_{\rm eq,2}/\eta_{\rm eq,1}=500$.  We contrast these to a selection of future GW experiments: LISA (blue) with an observing run of four years~\cite{Caprini:2019egz}, the Einstein Telescope (grey) for one year~\cite{Moore:2014lga,Maggiore:2019uih,Inomata:2018epa}, the Cosmic Explorer~\cite{Reitze:2019iox} (pink), DECIGO (brown) with three units for one year~\cite{Moore:2014lga,Kawamura:2020pcg}, $\mu$Ares (purple) for ten years~\cite{Sesana:2019vho}, THEIA (cyan) for twenty years~\cite{thetheiacollaboration2017theia,Garcia-Bellido:2021zgu} and SKA~\cite{Janssen:2014dka} (green).}
    \label{fig:distribution_2}
\end{figure}

The above figures use a flat spectrum for the curvature spectrum, resulting in a GW spectrum that depends only on $k\eta_{\mathrm{eq},2}$.  However, in order to discuss potential experimental reach, it is important to include the observed spectral tilt.  This is done in Figs.~\ref{fig:Sensitivities_Monochromatic} and ~\ref{fig:distribution_2}.  The colors indicate the types of matter (PBH, delayed, and thick wall Q-balls).  In Fig.~\ref{fig:Sensitivities_Monochromatic} the type of curve (solid, dashed, dotted) indicates the length of the eMD epoch while in Fig.~\ref{fig:distribution_2} it indicates the width of the mass distribution.  

As expected, the PBH signal is generally larger than the delayed Q-balls which is generally larger than the thick wall Q-ball, although we see that for sufficiently large Gaussian widths the PBH signal can be suppressed to values beneath delayed Q-balls with sufficiently narrow Gaussian widths.  In these plots, we have set the reheating temperature and used \eqref{eq:reheat_time} to find the amplitude of the GW spectrum today, taking $g_*$ and $ g_{*s}$ to both be $106.75$ for $T_R\geq 100 \textrm{GeV}$ and $10.75$ for $T_R=1 \textrm{MeV}$.

We see that accounting for the finite length of the end of the eMD epoch makes it unlikely that upcoming experiments will significantly probe PBH and Q-ball-dominated epochs, although THEIA may slightly constraint a narrow range of PBH scenarios with sufficiently peaked mass spectra.  We observe that experimental improvements may push further into the PBH parameter space, but thick wall and delayed Q-balls are unlikely to be probed in the foreseeable future.  Shorter eMD epochs, which lead to smaller poltergeist peaks, are even further from the reach of upcoming experiments.

\section{Conclusion}
\label{sec:conclusion}

In this work, we have improved the calculation of the second-order GW signal produced by rapidly-ending early matter epochs by including the finite transition time between matter and radiation domination.  We have considered four physically-motivated scenarios: PBHs, delayed Q-balls, thick wall Q-balls, and thin wall Q-balls, and found that the they produce different signal profiles in the monochromatic limit.  In particular, the resonant peak is largest for PBHs (which have the fastest transition) and, contrary to expectations, is nearly absent for thin wall Q-balls even though the effective decay rate scales as a negative power of the mass.  We have demonstrated that a sufficiently detailed measurement of the GW spectrum can distinguish both the type of matter and the length of the early matter epoch in the monochromatic limit.

We then moved away from a monochromatic mass spectrum and considered log-normal mass distributions.  As expected, for sufficiently narrow distributions the monochromatic limit is recovered.  Broader mass distributions suppress the GW signal as expected.  Although the peak is suppressed, there is not a significant broadening of the signal as there is from shorter eMD epochs.  Thus, at least in principle, a sufficiently precise observation of the GW spectrum can determine the type of matter, the length of the eMD epoch, and the breadth of the mass distribution function of the matter.

Finally, we included the observed spectral tilt of the curvature spectrum and calculated the amplitude of the GW power spectrum $\Omega_{\mathrm{GW}} h^2$ to compare with upcoming experiments.  We see that the suppression due to the finite transition period between matter and radiation domination is unfortunately sufficient to prevent upcoming experiments from probing these scenarios except in very limited situations.

However, we restricted ourselves to considering $\eta_{\mathrm{eq},2} \slash \eta_{\mathrm{eq},1} \lesssim 10^3$, so that non-linear effects only impacted modes which entered during radiation domination and so have further suppression.  For longer eMD epochs, modes which enter at the beginning of matter domination, and thus set the scale of the poltergeist peak, become non-linear.  Ref.~\cite{Kawasaki:2023rfx} has recently developed some tools for exploring this non-linear regime.  While we leave this regime for future work, we note the possibility for these scenarios to produce larger signals that may be able to be probed experimentally.  We also did not include the effect of non-Gaussianities, which can modify the signal~\cite{He:2024luf,Papanikolaou:2024kjb}.  We also note, though, that the optimistic conclusions in those papers should be tempered by the suppressions we present in this work.  

This paper is concerned primarily with the inherent suppression due to the different decay rates of different forms of matter and secondarily with suppression due to mass distributions.  Other effects may also further suppress the signal, such as the recently-discussed damping due to the mean free path of light particles~\cite{Domenech:2025bvr}.  As noted in that work, significant work remains to apply this to poltergeist-type signals due to the necessity of calculating on the resonance, and so also leave its inclusion for future work.  This is particularly true for Q-ball models, in which beyond-the-Standard-Model physics may be relevant.  We note, however, that Ref.~\cite{Domenech:2025bvr} indicates it is small when the early matter dominated epoch ends at temperatures above the electroweak scale, as on the right sides of Fig.~\ref{fig:Sensitivities_Monochromatic} and Fig.~\ref{fig:distribution_2}.

\vspace{2mm}
MP was supported by an Australian Government Research Training Program (RTP) scholarship and a Monash Graduate Excellence scholarship (MGES).  CB acknowledges support from the Australian Research Council via projects DP210101636, DP220100643, and LE210100015.  GW acknowledge the STFC Consolidated Grant
ST/X000583/1.

\bibliography{references}

\appendix
\section{Evolving the Background}
\label{ap:background}

In this appendix, we discuss some technical details of how we solved the background equations \eqref{eq:fried_m}, \eqref{eq:fried_r}, \eqref{eq:fried_a} with the decay rates derived in section \ref{sec:decay_rates}.  First we present a physical time analysis which allows for some analytical results and is mathematically simpler.  However, conformal time was used in presenting the results in References ~\cite{Inomata:pbh,Papanikolaou:2022chm,White:2021hwi,Kasuya:2022cko}, which made the instantaneous transition approximation.  Therefore, in order to compare our results to theirs, it is also convenient to work in conformal time, and we discuss how we implemented this approach.

As noted above, the decay rates are most conveniently expressed in physical time, in contrast to the interpolation tanh profile in conformal time used in \cite{White:2021hwi}.  Therefore, we transform the background equations to physical time,
\begin{align}
\dfrac{\partial \rho_m}{\partial t} &= - 3 \dfrac{\dot{a}}{a}  \rho_m - \Gamma(t) \rho_m, \nonumber \\
\dfrac{\partial \rho_r}{\partial t} &= -4 \dfrac{\dot{a}}{a} \rho_r + \Gamma(t) \rho_m,  \nonumber \\
\label{eq:backgrounds_physicalt}
\dfrac{\dot{a}}{a} &= \dfrac{\sqrt{\rho_{\mathrm{tot}}}}{\sqrt{3} M_{\mathrm{Pl}}} ,
\end{align}
where a dot denotes differentiation with respect to physical time.  It is possible to partially decouple these equations, by expressing the matter density as:
\begin{align}
\rho_m(t) &= \rho_{m,i} f(t) \left( \dfrac{a_0}{a(t)} \right)^3
 \label{eq:rho_m_remove_dilution}
\end{align}
where $f(0)=1$.  If the matter did not decay, we would have $f(t) = 1$ at all times and matter would just dilute with the expansion of the universe.  More generally, this function obeys the differential equation
 \begin{align}
 \dot{f}(t) &=  - \Gamma(t)  f(t),
\end{align}
which for decay rates of the form of eq.~\eqref{eq:gendecayrate} has the solution 
\begin{align}
f(t) &= \left( 1 - \dfrac{t }{t_{\mathrm{eva}}} \right)^{n}, 
\end{align}
which is valid for $t \leq t_{\mathrm{eva}}$. Therefore the energy density in matter is 
\begin{align}
\rho_m(t) &= \rho_{m,i} \left( \dfrac{a_0}{a(t)} \right)^3
\left( 1 - \dfrac{t }{t_{\mathrm{eva}}} \right)^{n}
\Theta\left(  t_{\mathrm{eva}} -t \right) .
\end{align}

However, the other two equations remain coupled and do not appear to have analytic solutions.  Expressing the radiation energy density as
\begin{align}
\rho_r(t) &= \rho_{r,i} g(t) \left( \dfrac{a_0}{a(t)} \right)^4,
\end{align}
we find its equation is 
\begin{align}
 \dot{g}(t) 
&=
 \Gamma(t) \dfrac{\rho_{m,i}}{\rho_{r,i}} \dfrac{a(t)}{a_0} \left( 1 - \dfrac{t }{t_{\mathrm{eva}}} \right)^{n} \Theta\left(  t_{\mathrm{eva}} -t \right) ,
\end{align}
with the initial condition that $g(0)=1$.  The scale factor equation is
\begin{align}
\dfrac{\dot{a}(t)}{a(t)} &= \dfrac{1}{\sqrt{3} M_{\mathrm{Pl}}} 
\sqrt{ 
 \rho_{r,i} g(t) \left( \dfrac{a_0}{a(t)} \right)^4
+  \rho_{m,i} \left( \dfrac{a_0}{a(t)} \right)^3
\left( 1 - \dfrac{t }{t_{\mathrm{eva}}} \right)^{n} \Theta\left(  t_{\mathrm{eva}} -t \right) 
}.
\end{align}
where one must choose a time at which $a(t) = a_0$ as an ``initial'' condition.  To solve these, it is necessary to also specify the initial ratio of matter to radiation energy density, $\rho_{m,i} \slash \rho_{r,i}$, as well as the ratio $t_{\mathrm{eva}} \sqrt{ \rho_{r,i}} \slash M_{\mathrm{Pl}}$.  As explained below, one is set by requiring  $\lim_{t \rightarrow 0} a(t) = 0$, consistent with radiation domination prior to matter domination.  The other is adjusted to determine the length of the eMD epoch.  Once these have been set, we can solve the remaining two equations numerically, once a time has been chosen for the scale factor condition.  

When we consider non-trivial mass distributions, we make the replacement 
\begin{align} \Gamma(t) \rho_m  \rightarrow  \rho_{m,i} \int \Gamma(t_{\mathrm{eva}},t) \beta(t_{\mathrm{eva}},t) \, d\ln t_{\mathrm{eva}},
\end{align}
where we note that $t_{\mathrm{eva}}$ describes the evaporation time of some initial mass $M_i$ (which, for Q-balls, may be expressed in terms of the initial charge $Q_i$).   The derivation proceeds similar to above, although we make the substitution
\begin{align}
\beta(X,t) &= f(X,t) \left( \dfrac{a_i}{a(t)} \right)^3
\end{align}
in place of \eqref{eq:rho_m_remove_dilution}.  We arrive at the same solution for the function $f(t)$ and therefore the matter energy density distribution function is
\begin{align}
\label{eq:distributiontdependence}
\beta(t_{\mathrm{eva}},t) &= \beta(t_{\mathrm{eva}},0)
\left( \dfrac{a_i}{a(t)} \right)^3 \left(1 - \dfrac{t}{t_{\mathrm{eva}}} \right)^{n} 
 \Theta(t-t_{\mathrm{eva}})
\end{align}
and we remember that the distribution variable $t_{\mathrm{eva}}$ gets integrated over.  The generalization of the other two equations is straightforward.

As mentioned, earlier results that made the instantaneous transition approximation presented their results using conformal time, and therefore although the above analysis is more mathematically straightforward we chose to work with conformal time in our code, which we now describe.  This is rather more mathematically and computationally complex, as the decay rate $\Gamma$ is expressed in physical time and furthermore, since the physical and conformal times are connected by the scale factor, the matter equation no longer fully decouples.

We rescale the energy densities as 
\begin{align}
x_m &= \dfrac{a^3}{a_{\mathrm{eq},1}^3} \dfrac{\rho_m}{\rho_{\mathrm{eq},1}} , \qquad 
x_r = \dfrac{a^4}{a_{\mathrm{eq},1}^4} \dfrac{\rho_r}{\rho_{\mathrm{eq},1}} 
\end{align}
where as above, the subscript $\mathrm{eq},1$ indicates the initial matter-radiation equality, which is the start of the eMD epoch.  Using a prime to indicate differentiation with respect to the unitless variable $\bar{\eta} = \eta \slash \eta_{\mathrm{eq},1}$, these satisfy:
\begin{align}
x_m^\prime &= - \bar{a} \bar{\Gamma} x_m \nonumber \\
x_r^\prime &= \bar{a}^2 \bar{\Gamma} x_m
\end{align}
where $\bar{a} = a \slash a_{\mathrm{eq},1}$ and $\bar{\Gamma} = a_{\mathrm{eq},1} \eta_{\mathrm{eq},1} \Gamma(t)$.  The scale factor equation becomes
\begin{align}
\bar{a}^\prime &= \alpha \sqrt{\bar{a} x_m + x_r}  
\end{align}
where 
\begin{align}
\alpha &= \dfrac{a_{\mathrm{eq},1} \eta_{\mathrm{eq},1} \sqrt{ \rho_{\mathrm{eq},1}}}{\sqrt{3} M_{\mathrm{Pl}}}.
\end{align}
Because the decay rates are expressed in terms of physical time, we need one more equation, which arises from $\eta = \int_0^t (1 \slash a(\tilde{t})) \, d\tilde{t}  $.  This equation becomes
\begin{align}
\bar{t}^\prime &= \bar{a},
 \end{align}
where $\bar{t} = t \slash (a_{\mathrm{eq},1} \eta_{\mathrm{eq},1})$.  This set of four differential equations can be solved numerically once parameters and initial conditions are set.  We note that by definition, at $\eta = \eta_{\mathrm{eq},1}$, $\bar{a} = 1$, $x_r = 1$, and $x_m = 1$, which are mathematically sufficient to serve as initial conditions (although in practice we begin our numerical analysis in the RD epoch before the eMD epoch).  These are equivalent to our initial conditions $g(0) = 1$, $f(0) = 1$, and $a(t) = a_0$ above, with the choice $t = t_{\mathrm{eq},1}$.  

As above, two constants remain undetermined (here, $\alpha$ and $\bar{t}_{\mathrm{eva}} = t_{\mathrm{eva}} \slash (a_{\mathrm{eq},1} \eta_{\mathrm{eq},1})$).  As above, we fix one by requiring $\lim_{\eta \rightarrow 0} a(\eta) = 0$.  If we consider times early enough that the decay rate is irrelevant, then $x_m = 1$ and $x_r = 1$, and the scale factor satisfies the equation 
\begin{align}
\bar{a}^\prime &= \alpha \sqrt{\bar{a} + 1}
\end{align}
which has the solution
\begin{align}
\bar{a}(\eta) &= \dfrac{1}{4} \left( 
\alpha^2 \eta^2 + 2 \alpha \eta C + C^2 -4 
\right) 
\end{align}
with $C$ an undetermined constant of integration.  Imposing $\bar{a}(0) = 0$ sets this constant to $\pm 2$.  If we assume that the decay is negligible at the start of the eMD epoch, then we can use this expression for the scale factor in the constraint $\bar{a}(1) = 1$.  Taking $C = 2$, this sets $\alpha = 2 (-1 \pm \sqrt{2}) $, whereas taking $C = -2$, this sets $\alpha = 2 (1 \pm \sqrt{2})$.  Requiring $a(\eta)$ to always be positive selects two possibilities, both of which simplify to
\begin{align}
a(\eta) &= \left(\sqrt{2}-1\right) \eta  \left(\left(\sqrt{2}-1\right) \eta +2\right).
\end{align}
We use $\alpha = 2 (-1 + \sqrt{2})$ (which corresponds to $C= -2$) in solving the coupled differential equations at later times, when the decay rate is not negligible.

\section{Effective Decay Rates of Log-Normal Distributions}
\label{ap:Decay}
In appendix~\ref{ap:background} we derived the time dependence of the mass distributions $\beta$ in eq.~\eqref{eq:distributiontdependence}.  In this appendix we describe how we treat the extended mass distribution throughout the evolution of the background and perturbations.  From the physical time continuity equation for the matter energy density in eq.~\eqref{eq:backgrounds_physicalt}, we can rearrange for $\Gamma(t)$ to define an effective decay rate
\begin{equation}
	\label{eq:effective_decay}
    \Gamma_{\rm eff}=-\frac{\dot{\rho}_m}{\rho_m}-3H.
\end{equation}

For a log-normal initial mass distribution, as in eq.~\eqref{eq:initial_distribution}, accounting for the evolution with time described in eq.~\eqref{eq:distributiontdependence}, the full time dependent distribution is given by
\begin{equation}
    \beta(t_{\rm eva},t)=\frac{\mathcal{N}}{\sqrt{2\pi}\sigma}\exp\left(-\frac{n^2\ln(t_{\rm eva}/t_{\rm eva,0})^2}{2\sigma^2}\right)\left(1-\frac{t}{t_{\rm eva}}\right)^n\left(\frac{a_i}{a(t)}\right)^3.
\end{equation}
To simplify the following analysis, we introduce the dimensionless variables
\begin{equation}
	x = \frac{t}{t_{\rm eva,0}},\quad \lambda = \ln\left(\frac{t_{\rm eva}}{t_{\rm eva,0}}\right),
\end{equation}
which allow the matter energy density to be expressed as
\begin{equation}
	\rho_{m}(x)=\rho_{m,i}\left(\frac{a_i}{a(x)}\right)^3\frac{\mathcal{N}}{\sqrt{2\pi}\sigma}\int_{\ln(x)}^{\infty}\left(1-x e^{-\lambda}\right)^n\exp\left(-\frac{n^2\lambda^2}{2\sigma^2}\right)d\lambda,
\end{equation}
where the lower integration bound accounts for objects that have completely evaporated.  For the decay rate, we also require the time derivative
\begin{equation}
    \dot{\rho}_m=-3H\rho_m-\rho_{m,i}\left(\frac{a_i}{a(x)}\right)^3\frac{\mathcal{N}}{\sqrt{2\pi}\sigma}\frac{n}{t_{\rm eva,0}}\int_{\ln(x)}^{\infty}e^{-\lambda}\left(1-x e^{-\lambda}\right)^{n-1}\exp\left(-\frac{n^2\lambda^2}{2\sigma^2}\right)d\lambda,
\end{equation}
which, after integrating by parts, reduces to
\begin{equation}
	\dot{\rho}_m=-3H\rho_m-\rho_{m,i}\left(\frac{a_i}{a(x)}\right)^3\frac{\mathcal{N}}{\sqrt{2\pi}\sigma}\frac{n^2}{\sigma^2 x t_{\rm eva,0}}\int_{\ln(x)}^{\infty}\lambda\left(1-x e^{-\lambda}\right)^n\exp\left(-\frac{n^2\lambda^2}{2\sigma^2}\right)d\lambda.
\end{equation}
By defining the dimensionless integrals
\begin{align}
	I_0(x)&=\int_{\ln(x)}^{\infty}\left(1-x e^{-\lambda}\right)^n\exp\left(-\frac{n^2\lambda^2}{2\sigma^2}\right)d\lambda, \\
	I_1(x)&=\int_{\ln(x)}^{\infty}\lambda\left(1-x e^{-\lambda}\right)^n\exp\left(-\frac{n^2\lambda^2}{2\sigma^2}\right)d\lambda,
\end{align}
we can express the effective decay rate simply as
\begin{equation}
	\Gamma_{\rm eff}(t) = \frac{n^2}{\sigma^2 t}\frac{I_1(x)}{I_0(x)}.
\end{equation}
Since the decay rate has to be evaluated many times in the evolution of the backgrounds and perturbations, we numerically compute $\Gamma_{\rm eff}$ for a grid of points and interpolate.  To improve the accuracy of the interpolation, we also compute the derivative
\begin{equation}
	\frac{d\Gamma_{\rm eff}}{d t} = \frac{n^2}{\sigma^2 t^2}\left(-\frac{I_1}{I_0}+\frac{I_2}{I_0}-\frac{n^2}{\sigma^2}\frac{I_1^2}{I_0^2}\right),
\end{equation}
where
\begin{equation}
	I_2 = \int_{\ln(x)}^{\infty}\left(\frac{n^2}{\sigma^2}\lambda^2-1\right)\left(1-x e^{-\lambda}\right)^n\exp\left(-\frac{n^2\lambda^2}{2\sigma^2}\right)d\lambda.
\end{equation}
To accurately compute the above integrals, both for small and large $x$, careful analysis is required.

For $x\ll 1$, $I_0$ is almost entirely dominated by the Gaussian term.  The effect from the decay term is negligible, except near the lower integration limit $\lambda=\ln(x)$, where the contribution to the integral is extremely suppressed by the Gaussian term.  Therefore, a good analytic approximation in this regime can be obtained from a binomial expansion of the first term.
\begin{align}
	I_0(x)&\simeq \sum_{i=0}^{m}\binom{n}{i}(-x)^i\int_{\ln(x)}^{\infty} e^{-i\lambda}\exp\left(-\frac{n^2\lambda^2}{2\sigma^2}\right)d\lambda \\
	&=\sqrt{\frac{\pi}{2}}\frac{\sigma}{n}\sum_{i=0}^{m}\binom{n}{i}(-x)^i\exp\left(\frac{i^2\sigma^2}{2 n^2}\right)\textrm{erfc}\left(\frac{n \ln(x)}{\sqrt{2}\sigma}+\frac{i\sigma}{\sqrt{2}n}\right),
\end{align}
where $\textrm{erfc}$ is the complimentary error function.  If $n$ is an integer, this provides an exact expression for the matter energy density and decay rate for $m=n$.  We use this with $m=2$ to get an accurate expression for both $\rho_m$ and its first derivative.  We can then write
\begin{equation}
	\rho_m(x) = \rho_{m,i}\left(\frac{a_i}{a(x)}\right)^3\frac{\mathcal{N}}{\sqrt{2\pi}\sigma}I_0(x),
\end{equation}
and compute the decay rate using eq.~\eqref{eq:effective_decay}.

The analytic approximation quickly breaks down, as the error in the binomial expansion at $\lambda=\ln(x)$ is large.  In this regime we switch to numerical evaluation.  However, for $x<1$, we still run into issues as the influence of the decay term can be washed out by the suppression from the Gaussian.  This is exacerbated in regions where $x$ is still small, as $I_1$ and $I_2$ evaluate to approximately zero, with cancellations during the numerical evaluation leading to loss of precision, and the effect of the decay term is completely lost.  This can be rectified by observing approximate symmetries present in the integrands; even for $I_0$ and $I_2$ and odd for $I_1$.  Therefore, we can exploit these symmetries to perform these cancellations analytically
\begin{align}
	I_0(x)=&\int_{ln(x)}^{0}\left[\left(1-x e^{-\lambda}\right)^n+\left(1-x e^{\lambda}\right)^n\right]\exp\left(-\frac{n^2\lambda^2}{2\sigma^2}\right)d\lambda\nonumber\\
	&+\int_{-ln(x)}^{\infty}\left(1-x e^{-\lambda}\right)^n\exp\left(-\frac{n^2\lambda^2}{2\sigma^2}\right)d\lambda, \\
	I_1(x)=&\int_{ln(x)}^{0}\left[\left(1-x e^{-\lambda}\right)^n-\left(1-x e^{\lambda}\right)^n\right]\lambda\exp\left(-\frac{n^2\lambda^2}{2\sigma^2}\right)d\lambda\nonumber\\
	&+\int_{-ln(x)}^{\infty}\left(1-x e^{-\lambda}\right)^n\lambda\exp\left(-\frac{n^2\lambda^2}{2\sigma^2}\right)d\lambda, \\
	I_2(x)=&\int_{ln(x)}^{0}\left[\left(1-x e^{-\lambda}\right)^n+\left(1-x e^{\lambda}\right)^n\right]\left(\frac{n^2}{\sigma^2}\lambda^2-1\right)\exp\left(-\frac{n^2\lambda^2}{2\sigma^2}\right)d\lambda\nonumber\\
	&+\int_{-ln(x)}^{\infty}\left(1-x e^{-\lambda}\right)^n\left(\frac{n^2}{\sigma^2}\lambda^2-1\right)\exp\left(-\frac{n^2\lambda^2}{2\sigma^2}\right)d\lambda  .
\end{align}
After making this decomposition, the above integrals are all straightforward to evaluate numerically.

Lastly, for $x\gg 1$, we are unable to evaluate the integrals accurately due to underflow from the Gaussian terms.  Therefore for $x\geq 1$ we transform to the integration variable
\begin{equation}
	u = \lambda -\ln(x),
\end{equation}
resulting in the integrals
\begin{align}
	I_0(x)&=A\int_{0}^{\infty}\left(1- e^{-u}\right)^n\exp\left(-\frac{n^2}{2\sigma^2}\left(u^2+2\ln(x)u\right)\right)d\lambda, \\
	I_1(x)&=A\int_{0}^{\infty}(u+\ln(x))\left(1- e^{-u}\right)^n\exp\left(-\frac{n^2}{2\sigma^2}\left(u^2+2\ln(x)u\right)\right)d\lambda, \\
	I_2(x)&=A\int_{0}^{\infty}\left(\frac{n^2}{\sigma^2}(u+\ln(x))^2-1\right)\left(1- e^{-u}\right)^n\exp\left(-\frac{n^2}{2\sigma^2}\left(u^2+2\ln(x)u\right)\right)d\lambda,
\end{align}
where
\begin{equation}
	A=\exp\left(-\frac{n^2\ln(x)^2}{2\sigma^2}\right).
\end{equation}
The underflow is absorbed into the prefactor $A$, which cancels from the evaluation of the decay rate, as $\Gamma_{\rm eff}$ and its derivative are expressed only in terms of the ratios of the various integrals.

\end{document}